\documentclass{aastex} 
\usepackage{emulateapj5, apjfonts, psfig, epsfig}
\begin{document}
\received{}
\accepted{}
\revised{}
\slugcomment{The Astronomical Journal, in press}
\shortauthors{Corwin et al.}
\shorttitle{M75. II. RR Lyrae Variables}

\title{M75, a Globular Cluster with a Trimodal Horizontal Branch.\\
   II. $BV$ photometry of the RR Lyrae Variables}

  \author{T.~M.~Corwin\altaffilmark{1}}
  \affil{Department of Physics, University of North Carolina at Charlotte, 
Charlotte, NC 28223\\email: mcorwin@uncc.edu} 

  \author{M.~Catelan}
   \affil{Pontificia Universidad Cat\'olica de Chile, Departamento de 
Astronom\'\i a y Astrof\'\i sica, \\ Av. Vicu\~{n}a Mackenna 4860, 
782-0436 Macul, Santiago, Chile \\email: mcatelan@astro.puc.cl} 

  \author{H.~A.~Smith}
  \affil{Department of Physics and Astronomy, Michigan State University, East 
Lansing, MI 48824-2320\\email: smith@pa.msu.edu}  

  \author{J.~Borissova}
  \affil{Institute of Astronomy, Bulgarian Academy of Sciences, 
  72~Tsarigradsko chauss\`ee, BG\,--\,1784 
  Sofia, Bulgaria\\and\\
  Pontificia Universidad Cat\'olica de Chile, Departamento de 
  Astronom\'\i a y Astrof\'\i sica, \\ Av. Vicu\~{n}a Mackenna 4860, 
  782-0436 Macul, Santiago, Chile\\email: jborisso@astro.puc.cl} 

  \author{F.~R.~Ferraro}
  \affil{Osservatorio Astronomico di Bologna, Via Ranzani 1, I-40127 Bologna, 
Italy\\email: ferraro@apache.bo.astro.it} 

  \and 
  \author{W.~S.~Raburn}
  \affil{Department of Physics, University of North Carolina at Charlotte, 
Charlotte, NC 28223\\email: wsraburn@uncc.edu}

\altaffiltext{1}{Visiting Astronomer, Cerro Tololo Inter-American Observatory, 
National Optical Astronomy Observatories, which are operated by AURA, Inc., 
under cooperative agreement with the National Science Foundation.} 

\begin{abstract}
We present new $BV$ CCD photometry, light curves, and ephemerides for 9 
previously known, 29 newly detected RR Lyrae variables and 
one newly detected variable of an unknown type in the globular 
cluster M75. The photometry used for the detection of the additional variables 
was obtained with the image subtraction package {\sc isis}. The data were acquired 
on an observing run in July 1999 and range over 7 observing nights. Estimates of 
fundamental photometric parameters are presented including intensity- and 
magnitude-averaged $B$ and $V$ magnitudes, magnitude-averaged colors, pulsation 
periods, and pulsation amplitudes. The mean period of the RRab variables, 
$\langle P_{\rm ab} \rangle = 0.5868$~d, and the number fraction of RRc stars, 
$N_{\rm c}/N_{\rm RR} = 0.342$, are both large for an Oosterhoff type I (OoI) 
globular cluster, suggesting that M75 may be Oosterhoff-intermediate. 
Possible conflicts between Oosterhoff-type determination based on the 
$A_V - \log\,P$ and $A_B - \log\,P$ diagrams are discussed. 
The physical parameters of the RRc and RRab variables, as obtained from Fourier  
decomposition of their light curves, do not show any clear deviation from 
normal OoI behavior. 
\end{abstract}


\section{Introduction}
M75 (NGC 6864) is a distant globular cluster ($R_{\odot} \simeq 19$~kpc; Harris 
1996) lying on the other side of the galactic center. The cluster has a 
relatively high degree of concentration 
($\log[\rho_0/(M_{\odot}{\rm pc}^{-3})] = 4.9$; 
Pryor \& Meylan 1993). Catelan et al. (2002, hereafter Paper~I) found a 
metallicity of [Fe/H] = $-1.03 \pm 0.17$~dex and $-1.24 \pm 0.21$~dex in the 
Carretta \& Gratton (1997) and Zinn \& West (1984) scales respectively. Its 
reddening, $E(\bv)$, is $0.16 \pm 0.02$~mag (Paper~I). 
 
M75 has been the target of relatively few studies. Harris (1975) provided the 
first color-magnitude diagram (CMD). Paper~I provided a CMD based on CCD data 
and showed that M75, suspected to have a bimodal horizontal branch (HB) 
by Catelan et 
al. (1998), actually has a trimodal HB. The focus of this paper, Paper~II in the 
series, will be the RR Lyrae variables in M75.

RR Lyrae variables provide crucial information for estimating globular cluster 
ages and distances, as summarized by Smith (1995). They are easily identified by 
their distinctive light curves and are bright enough to be observed to 
considerable distances. Their absolute magnitudes appear to be quite restricted. 
The range of RR Lyrae luminosities is discussed in Carney (2001) and Harris 
(2001).   

As we have shown in Paper~I, M75 has a special HB morphology: it is one among 
only a handful of globular clusters with a bimodal---actually a 
{\em tri}modal---distribution 
of HB stars. The physical cause of HB bimodality is still not 
known (Rood et al. 1993; Catelan et al. 1998). RR Lyrae variables can provide 
precious information on the ``second parameter(s)" that lead to HB bimodality 
(Borissova, Catelan, \& Valchev 2001 and references therein), since different 
second parameter candidates are expected to affect RR Lyrae luminosities, masses 
and pulsational properties in different ways (see, e.g., Catelan 1996). For this 
reason, variability surveys of bimodal-HB globulars which contain sizeable RR 
Lyrae populations, such as NGC~1851 and M75, are of great potential importance 
in sheding light on the connection between bimodal HBs and the second-parameter 
effect. Our interest in these variables was furthered by the confirmation, in
Paper~I, of the suggestion by Harris (1975) 
that M75 seems to have a high number ratio $R$ of HB stars to red giant branch 
(RGB) stars, which could be explained by a higher-than-standard helium abundance 
$Y$ (e.g., Iben 1968). If so, the M75 HB stars should be brighter than those in 
other clusters with ``normal'' $R$, implying longer periods for their RR Lyrae 
variables.  

The variables in M75 were previously studied by Pinto, Rosino, \& Clement (1982, 
hereafter PRC82). Their study revealed ten RR Lyrae variables, although $B$ 
light curves were constructed for only six of them. Several hints existed that 
the PRC82 study was incomplete, thus making a new variability survey necessary: 
(i) M75 is a distant and dense cluster; (ii) PRC82 only obtained photographic 
photometry; (iii) Recent results (Kaluzny, Olech, \& Stanek 2001) using the new 
image-subtraction techniques (Alard 2000; Alard \& Lupton 1998) suggest that 
even nearby, well-studied globular clusters---particularly the more concentrated
ones---suffer from significant 
incompleteness in their known RR Lyrae populations.  The purpose of this paper 
is to present the first extensive CCD variability survey for M75.

\section{Observations and Reductions}
\subsection{Observations}

The CCD images used in this study were obtained with the 0.9-meter telescope at 
the Cerro Tololo Inter-American Observatory. The field was observed over a seven 
night interval in July 1999. Observing conditions were not good for three of the 
seven nights and data from these nights are not included here. The data reported 
here were obtained on the nights of 15/16, 19/20, 20/21, and 21/22 July 1999. 
The 2048$\times$2048 Tek2K-3 CCD was used. Typical exposure times were 240 s for 
the $V$ frames and 360 s for the $B$ frames. The pixel scale was 0.395 arcsec 
giving a field of view of 13.5 arcmin. 

\subsection{Data Reductions}

The raw data frames were processed following standard procedures to remove the 
bias, trim the pictures, and divide by mean dome flats obtained using 
color-balanced filters. No attempt was made to recover bad pixels or columns. 
Initially raw magnitudes were derived using {\sc allstar} and {\sc daophot} 
(Stetson 1994). 
In this reduction only 7 variables, all priviously identified, were clearly 
present. Next we applied {\sc allframe} (Stetson 1994) to the images. The results 
yielded light curves for 9 of the previously identified variables and resulted 
in the discovery of 13 additional variables. 

As a final attempt to improve the quality of the light curves and perhaps detect 
additional variables, we employed the image subtraction package {\sc isis} V2.1 (Alard 
\& Lupton 1998; Alard 2000). The resulting differential flux data produced 
improved light curves for all of the variables found using {\sc allframe} and found an 
additional 17 variables, giving 39 variables in all. These results are 
consistent with the predictions by Kaluzny et al. (2001) of a greater than 30\% 
incompleteness factor in the detection of RR Lyrae variables in globular 
clusters.\footnote{We cannot provide a finding chart identifying the new variables
because they are in the crowded cluster center, and we do not have an exposure 
that shows individual stars there.}

In order to convert the {\sc isis} differential flux data to standard magnitudes we 
used {\sc allstar} and {\sc daophot} (Stetson 1994) to obtain instrumental magnitudes for 
each of the variables in the $B$ and $V$ reference images of the {\sc isis} 
reductions. Due to the distance of the cluster and crowding in its center, the 
photometry for many of the variables was unreliable. We choose a $\chi$ value of 
3.5 as the dividing line between reliable and unreliable photometry. With this 
criterion, 14 variables had reliable photometry in both the $B$ and $V$ filters, 
13 had reliable photometry in $B$ but not $V$, and 2 had reliable photometry in 
$V$ but not $B$. 

For those variables with {\sc allstar} $\chi$ values of less than 3.5, the instrumental 
magnitude for each variable was converted to flux. The {\sc isis} differential flux 
was then added to the reference flux to determine the total flux for each 
variable in each of the exposures, and these fluxes were converted back to 
instrumental magnitudes. Using the standard magnitude data from Paper~I, 
transformation equations were derived to convert the instrumental magnitudes 
from the {\sc isis} reductions into standard magnitudes. The 14 variables with $\chi$ 
less than 3.5 in both filters were transformed to the standard system using 
transformation equations of the following form:
\begin{equation}
V-v_{0} = \epsilon(b-v)_{0} + \zeta_{v}
\end{equation}
\begin{equation}
(\bv) = \mu(b-v)_{0} + \zeta_{bv}.
\end{equation}

The 15 variables with good photometry in only one of the filters were also put 
on the standard system for the appropriate filter. An appropriate value for the 
color (depending on the mode of pulsation) was assumed. That is, a value of 
instrumental $b-v$ was assumed that would give $\bv = 0.35$ mag for RRc 
variables and $\bv = 0.50$ mag for RRab variables. In this way standard $B$ 
magnitudes were computed for the 13 variables with good $B$ and bad $V$ 
instrumental magnitudes, and standard $V$ magnitudes were computed for the two 
variables with good $V$ and bad $B$ instrumental magnitudes. There are two 
sources of error associated with this. As an RR Lyrae pulsates its color changes 
in proportion to its amplitude, with $\Delta(\bv)$ having a maximum of about 
0.4~mag. This effect will increase the $B$ amplitude of pulsation listed in 
Table~1 for the 13 stars with good $B$ and bad $V$ photometry by a maximum 
of about 0.044~mag (roughly a 0.02~mag extension at both the maximum and 
minimum brightness). This error in amplitude decreases with amplitude, being 
less than 0.022~mag for RRc variables. The color change over the pulsation 
cycle has very little effect on the average $B$ magnitudes. The second source 
of error involves assuming a particular value of $\bv$ for the RRab and RRc 
variables. The actual range of magnitude-averaged colors for each of these is 
about 0.12~mag. This could affect the average $B$ magnitudes by $\pm 0.007$~mag 
relative to the values listed in Table~1. Because of the form of the transformation 
equations, there is negligible effect on the values for $V$ amplitudes and magnitudes 
listed in Table~1 for the two stars with good $V$ and bad $B$ photometry.

\section{Periods and Light Curves}
Periods were determined using the period-search program ``kiwi.'' (``Kiwi'' 
searches for periodicity by seeking to minimize the total length of the line 
segments that join adjacent observations in phase space, i.e., to maximize the 
smoothness of the light curve. The ``kiwi'' program was kindly provided to us by 
Dr. Betty Blanco.) 

The location of the variables in arcsec from the center of the cluster as in 
Sawyer-Hogg's (1973) catalog, and the new periods and ephemerides are given in 
Table 1. The data cover 7 days, spanning about 21 cycles for the shorter period 
variables and only about 9 cycles for the longer period variables. Because only 
the current data are used to determine periods, the periods are quoted to only 
four significant figures. Light curves based on the periods given in Table 1 are 
shown in Figure 1. For the variables with good photometry in both filters, the 
light curves are in standard magnitudes. For the other variables, the 
differential flux from the {\sc isis} reduction is used to produce the light curves. 
It is obvious that the photometry is of good internal precision, with the 
scatter about the mean loci being on the order of 0.02~mag. 

\section{Light Curve Parameters}

For each variable for which standard magnitudes were determined (either $B$, 
$V$, or both), the mean magnitudes have been derived by linear interpolation of 
the phase-magnitude diagrams. The mean values were derived both as magnitude-
weighted and as intensity-weighted means (Storm, Carney, \& Beck 1991). For the 
14 variables with standard magnitudes in both $B$ and $V$, the mean colors (the 
difference between magnitude-averaged $B$ and $V$) were determined. The $V$ and 
$B$ amplitudes were determined as the differences in the average maxima and 
minima of the light curves. These results are given in Table~1.

Figure~2 is a histogram of the M75 RR Lyrae periods with the RRab and RRc 
variables clearly separated. The 13 RRc variables all have periods less than 0.4 
day while the 25 RRab variables all have periods greater than 0.45 day. The 
average period for the RRab variables is 0.5868~day and for the RRc variables is 
0.3054 day. For its relatively high metallicity, one might expect M75 to be of 
Oosterhoff type I. However, the relatively long mean period of the M75 RRab 
variables would make it the Oosterhoff~I cluster with the longest value of 
$\langle P_{\rm ab}\rangle$, as can be seen from Figure~4 in Clement et al. (2001). 
Interestingly, the number ratio $N_{\rm c}/N_{\rm RR} = 0.342$ is also large 
for OoI standards, being intermediate between OoI and OoII values---Clement et 
al. list $N_{\rm c}/N_{\rm RR} = 0.22$ for OoI clusters, and  
$N_{\rm c}/N_{\rm RR} = 0.48$ for OoII clusters. Therefore, in terms of mean 
RRab periods and RRab-to-RRc number fraction alike, M75 might classify better 
as an Oosterhoff-intermediate cluster, rather than as an OoI cluster as would be 
expected for its metallicity. We will further test the possibility that
M75 is an Oosterhoff-intermediate cluster in \S6 
below, using the Fourier decomposition parameters of the M75 RR Lyrae light 
curves.  

If M75 is confirmed to be an Oosterhoff-intermediate globular, it is important
to note that there is one important property of M75 that 
distinguishes it from the the well-known Oo-intermediate globulars 
in the LMC (Bono, Caputo, \& Stellingwerf 1994): It is much more metal-rich than 
the latter, occupying a completely different position in the HB morphology-[Fe/H] 
plane (Fig.~3). To produce Figure~3, we have utilized data from Bono et al. for 
the LMC and OoI clusters (replacing the metallicity values and HB types for 
the latter with data from Harris 1996), and included, among the OoII clusters, 
in order of decreasing metallicity, M2 (NGC~7089), M53 (NGC~5024), M68 
(NGC~4590), NGC~5466, and M15 (NGC~7078). For these, the [Fe/H] and HB 
morphology values were also adopted from the Harris catalogue. This diagram 
shows that a different explanation might be required for M75's 
Oosterhoff-intermediate classification, if this is confirmed, than has been 
proposed for the LMC globulars (Bono et al. 1994). A tantalizing possibility 
is that such an Oosterhoff-intermediate classification could be related to 
M75's HB bimodality and peculiar $R$-ratio (Paper~I). On the other hand, it 
should be noted that there are also ``regular'' OoI clusters that present 
bimodal HBs, such as NGC~6229 (Borissova et al. 2001) and NGC~1851 (Walker 1998; 
but see also Bellazzini et al. 2001, especially their \S4.6).  

Figure~4 shows magnitude-averaged color as a function of period. Due to the 
strong temperature dependence implied by the period-mean density relation, it 
is expected that the longer periods correspond to the redder colors. NV8, NV3, 
and NV7 are somewhat anomalous in this plot, and will be discussed in further 
detail below. 

Figure~5 is a Bailey diagram, a plot of $B$ amplitude as a function of log 
period. As has long been known, for the RRab variables, the amplitude decreases 
as the period increases, while for the RRc variables, the amplitude first 
increases and then decreases with increasing period (Sandage 1981a). Also shown 
in the figure are ``typical'' lines for OoII and OoI clusters. The OoII line 
was obtained from C.~Clement's (priv. comm.) corresponding line in the 
$A_V - \log\,P$ 
diagram, transforming $V$ amplitudes to $B$ amplitudes by means of the relation 
$A_V = 0.72 \, A_B + 0.03$ (Layden et al. 1999). The ``OoI line'' is the same 
relation as provided by Borissova et al. (2001) for M3. 

The location of an RRab variable in the period-amplitude plane of Figure~5 is a 
measure of its average luminosity, which is determined by its surface 
temperature and radius (the Stefan-Boltzmann equation). Sandage (1981a, 1981b), 
Jones et al. (1992), Catelan (1998), and Sandquist (2000) have shown that the 
$B$ amplitude of an RRab variable is related to its effective temperature; the 
larger the amplitude, for a given metallicity, the higher the temperature (but 
see De Santis 2001). The pulsation equation relates the period of a variable 
star to its average density; the period is inversely proportional to the square 
root of the average density. Assuming that the masses of the variables are 
distributed in a narrow range, the period becomes a measure of the average size 
of the star; the longer the period, the larger the star. Thus, for a given 
amplitude (temperature), the longer period (larger) variables should be more 
luminous.

Assuming all of the variables are cluster members and thus at about the same 
distance, the position of a variable in the period-amplitude plane should 
correspond to its apparent brightness. There are, however, several anomalies. 
The three brightest RRab variables are NV23, NV24, and NV17 with 
intensity-averaged $B$ magnitudes of 16.960, 17.150, and 17.515~mag 
respectively. A higher 
luminosity for these stars is not consistent with their positions in the 
period-amplitude plane. In fact, NV23 with a short period (small size) and 
very low 
amplitude (low temperature) should have a relatively low luminosity. In fact, 
all three stars have small amplitudes for their periods. 

It is possible, of course, that for NV23, NV24, and NV17, one or more are not 
cluster members, and that their greater brightness results from their being 
closer than the cluster. This seems unlikely in that they are within 9, 12, 
and 30 arcsec, respectively, of the cluster center. 
Another possible explanation for the fact that they are 
brighter than their pulsation properties would suggest is that these stars are 
blends. This might account for the unusually small amplitudes of NV23 and NV17 
relative to what would be expected based on their periods. It should also be 
noted that the $D_{\rm m}$ parameter (Jurcsik \& Kov\'acs 1996), as discussed 
in more detail in \S6.2 below, also indicates all of these stars to be 
``anomalous.''

NV10 has a unique position in the period-amplitude plane. Its very long period 
(large size) relative to the other RRab variables of comparable amplitude 
(temperature) suggests a greater luminosity and hence brightness. In fact NV10 
does have an intensity-averaged $V$ magnitude of 17.432, significantly brighter 
than the ZAHB. As discussed by Clement \& Shelton (1999) and Lee, Demarque, \& 
Zinn (1990), a higher luminosity for NV10 might indicate that it is in an 
advanced evolutionary stage. For a recent, critical discussion of the possible 
evolutionary history of such bright RR Lyrae stars, the reader is referred to   
Pritzl et al. (2002).

For the RRab variables in Figure~5, there is considerable scatter. Clement \& 
Shelton (1999) pointed out that Blazhko variables introduce scatter into the 
period-amplitude diagram. The scatter is introduced when the light curve of a 
Blazhko 
variable represents it at less than its maximum amplitude. Thus yet another 
possible explanation for the low amplitudes of NV23 and NV17 (discussed 
above) is that they are Blazhko variables whose light curves represent them at 
much less than maximum amplitude. There is some evidence in the data that NV17 
is in fact a Blazhko variable.

For most Blazhko variables the difference in amplitude would not be apparent 
over an observing interval of just seven days. However, NV17 reached maximum 
brightness on three consecutive nights of the observing run, and the 
corresponding $B$ magnitudes are 16.925, 16.969, and 17.024 mag. Although NV17 
was not put on the standard system for $V$, the differential $V$ fluxes 
corresponding to maximum brightness for the three consecutive nights are $-
$8011.6, $-$6372.1, and $-$5917.7, showing the same pattern of decrease in 
maximum brightness. An additional test for Blazhko variables involves the 
Fourier decomposition parameters to be discussed in \S6.

\section{Color-Magnitude Diagram}

The CMD for M75 has been provided in Paper~I. Using these data, Figure~6 is a 
magnified CMD that highlights the HB and includes those variable 
stars for which we have $\bv$ values. In general, the variable stars fall in the 
appropriate areas of the instability strip according to their mode of pulsation. 
There are, however, two obvious anomalies. Based on its period, amplitude, and 
the shape of its light curve, NV3 is clearly an RRab variable. However, its 
color places it with the RRc variables on the CMD. A possible explanation is 
that it is blended with a bluer star. This is consistent with the fact that it 
is somewhat brighter than most other RR Lyrae variables and with the 
fact that it has an unusually low ratio of the $B$ to $V$ amplitudes (see Pritzl 
et al. 2002). 

The other anomaly is NV7 which lies well to the red side of the instability 
strip. Its intensity-averaged $V$ magnitude also shows it to be fairly bright, 
and it has a 
high ratio of the $B$ to $V$ amplitude, so it also might be a blended star, in 
this case with a redder companion. These stars are, of course, also discrepant 
in the color-period plot (Fig.~4). There is, however, no indication of 
blending in the {\sc allstar} round value for either NV3 or NV7. On the basis 
of their Fourier parameters (\S6.2), only NV7 is classified as ``anomalous.''

Another unusual feature of the CMD is that both NV10 and NV8 are considerably 
brighter than the ZAHB. Like NV3, NV8---an RRc star---has a low ratio of the 
$B$ to $V$ amplitudes and is probably blended with a bluer star.

\section{Physical Properties of the RR Lyrae Variables from their Fourier
Parameters}

Fourier decompositions of the RR Lyrae light curves were done fitting to an 
equation of the form: 

\begin{equation} 
mag = A_0 + \sum_{j=1}^{10} A_j \, \sin(j\omega t + \phi_j + \Phi),  
\end{equation} 

\noindent where $\omega = 2\pi/P$, and $\Phi = 0$ for the RRab, and 
$\Phi = \pi/2$ for the RRc (i.e., a 
sine decomposition was carried out for the RRab, but a cosine 
decomposition for the RRc). When the amplitude $A_j$ is found to be 
negative, we change its sign and add $\pi$ to the corresponding phase
$\phi_j$. 

\subsection{RRc Variables}

Amplitude ratios $A_{j1} = A_j/A_1$ and phase differences 
$\phi_{j1} = \phi_j - j \phi_1$ for the lower-order terms 
are provided in Table~2. Note that the phase 
differences are adjusted to a positive value between 0 and $2\pi$, which is 
accomplished by adding (or subtracting) multiples of $2\pi$ to (from) 
$\phi_{j1}$ as may be necessary. In this table (and in the ones to follow), 
uncertain values are indicated with a colon symbol (``:''), whereas unreliable 
ones are denoted by a double colon symbol (``::''), and are provided for 
completeness only. For NV29 and N30, whose RRc classification is uncertain 
(see Table~1 and Fig.~1), no measurements were attempted. The error in the 
$\phi_{31}$ coefficient was obtained from Eq.~(16d) of Petersen (1986). 

Simon \& Clement (1993) used light curves of RRc stars obtained from 
hydrodynamic pulsation models to derive equations to calculate mass $M$, 
luminosity $\log\,L$, temperature $T_{\rm e}$ and a ``helium parameter'' $y$ 
as a function of the Fourier phase 
difference $\phi_{31}$ and period. These correlations were based on light curves 
using magnitudes as a measure of variation in brightness. In their study 
of NGC 6229, Borissova et al. (2001) calculated physical parameters from the 
Simon \& Clement relationships using light curves based on both apparent 
magnitudes ({\sc daophot}, Stetson 1994) and differential fluxes ({\sc isis}). 
The fact that 
some of the {\sc isis} light curves were of better quality than the {\sc daophot} 
ones 
resulted in reduced errors for some of the Fourier parameters, but there were no 
significant differences between the two analyses in the average values of the 
physical parameters. 

Using Simon \& Clement's (1993) equations (2), (3), (6), and (7) and our apparent 
magnitude light curves, where available, and {\sc isis} light curves otherwise, 
we computed $M/M_{\odot}$, $\log\,(L/L_{\odot})$, $T_{\rm e}$ and $y$ for the RRc 
stars in M75. These results are given in Table~4. Note that the ``helium 
parameter'' $y$ is {\em not} supposed to provide a good description of the 
helium abundance (Simon \& Clement 1993), so that the deviations from canonical 
values for GC stars (i.e., $Y \approx 0.23-0.25$) that are apparent in this 
table should neither be considered reliable evidence for a high helium 
abundance in M75, nor necessarily a serious problem with the Simon \& Clement 
method. However, it should also be noted that similarly high $y$ values have 
previously been reported, e.g., by Borissova et al. (2001) and Olech et al. (2001). 
In like vein, several of the mass values given in this table, particularly 
those for V9, NV8, and NV12, are clearly too low, being 
lower than even the helium core mass at the He-flash (Catelan, de Freitas Pacheco, 
\& Horvath 1996 and references therein). 
Such low mass values are not uncommon in the literature either, 
having recently been reported, for example, by Borissova et al., Clement 
\& Rowe (2000), Olech et al. (2001), and Pritzl et al. (2001, 2002). In our opinion, 
these low RRc mass values are unphysical, and represent evidence that the Simon 
\& Clement method may indeed be affected by systematic errors. 

The unweighted mean values and corresponding standard errors of the mass, 
log-luminosity, effective temperature and ``helium parameter''  
are: $0.53 \pm 0.02 \, M_{\odot}$, $1.67 \pm 0.01$, $7399 \pm 37$~K and 
$0.289 \pm 0.003$, respectively.

\subsection{RRab Variables}

Amplitude ratios $A_{j1} = A_j/A_1$ and phase differences 
$\phi_{j1} = \phi_j - j \phi_1$ for the lower-order terms 
are provided in Table~3. Note that, as in the cos decompositions for the 
RRc's (\S6.1), the phase 
differences are adjusted to a positive value between 0 and $2\pi$, which is 
accomplished by adding (or subtracting) multiples of $2\pi$ to (from) 
$\phi_{j1}$ as may be necessary. In 
this table, the Jurcsik-Kov\'acs $D_{\rm m}$ value, which is intended to 
differentiate RRab stars with ``regular'' light curves from those with 
anomalies (e.g., the Blazhko effect), is also given (eighth column).  

Jurcsik \& Kov\'{a}cs (1996), Kov\'{a}cs \& Jurcsik (1996, 1997), 
Kov\'{a}cs \& Kanbur (1998), Jurcsik (1998) and Kov\'acs \& Walker (2001) 
obtained empirical formulae relating the stellar 
metallicities, absolute magnitudes and temperatures to Fourier decomposition 
parameters for RRab stars with ``regular'' light curves. The only model-dependent 
ingredients in their calibrations are the zero point of the HB luminosity scale 
(adopted from Baade-Wesselink studies) and the color-temperature transformations 
(obtained from static model atmospheres). Again, although this method was 
designed for use with apparent magnitude light curves in $V$, Borissova et al. 
(2001) 
have shown that the overall averages are little affected by using {\sc isis} 
relative
fluxes. As with the RRc variables, we have used a combination of apparent 
magnitude light curves and {\sc isis} light curves for our Fourier analysis 
of the RRab variables. 

The physical parameters of M75 RRab variables obtained from this method are 
given in Table~5. These values are determined using 5 RRab variables (V14, 
NV2, NV3, NV5, NV9) for which the parameter $D_{\rm m}$ is less than 
5.0. Variables with $D_{\rm m}$ less than 3.0 are designated by Jurcsik \& 
Kov\'{a}cs as regular, while those with larger $D_{\rm m}$ values are called 
``peculiar.'' The $D_{\rm m}$ values were obtained from Eq.~(6) and Table~6 
in Jurcsik \& Kov\'acs (1996); [Fe/H], $M_V$, $V-K$ and 
$\log\,T_{\rm e}^{\langle V-K \rangle}$ come from Eqs.~(1), (2), (5) and (11) 
of Jurcsik (1998). Eqs.~(6) and (9) of Kov\'acs \& Walker (2001) were used 
to compute the color indices $(\bv)$ and $(V-I)$, respectively; then Eq.~(12) 
of Kov\'acs \& Walker (1999) was used, assuming a mass of $0.7\,M_{\odot}$, 
to derive temperature values from Eq.~(11) (for $\bv$) and Eq.~(12) (for $V-I$) 
in Kov\'acs \& Walker (2001). 

For the criterion $D_{\rm m}$ less than 5.0 for regular variables, the unweighted 
mean value (and corresponding standard deviation) of [Fe/H] derived from 
$\phi_{31}$ is $-1.01 \pm 0.05$~dex. Note that this value is in the Jurcsik 
(1995) scale; this corresponds to $-1.32$~dex in the Zinn \& West (1984) 
scale (in reasonable agreement with the value found in Paper~I). 
Likewise, the mean absolute magnitude is $\langle M_V \rangle = 0.81 \pm 0.01$~mag. 
The faint HB is a reflection of the adoption of the Baade-Wesselink luminosity zero 
point in the calibration of this method (see Jurcsik \& Kov\'{a}cs 1999 for a recent 
discussion). The temperature values based on the $V-K$, $\bv$ and $V-I$ formulae are 
similar, the ones based on $\bv$ being higher than those based on the other color
indices by $\Delta\,\log\,T_{\rm e} \simeq 0.005$, or about 75~K.  

If the criterion for ``regular'' status is made more strict 
($D_{\rm m} < 3.0$), only V14 is removed from the list of ``regular'' 
variables, and the mean values of the physical parameters change only very 
slightly: The mean colors and temperatures become bluer (average change, 
considering the three temperature values: 
$\langle \Delta\,\log\,T_{\rm e} \rangle \simeq 0.0025$), 
the metallicity increases 
by $\Delta\,{\rm [Fe/H]} \simeq +0.04$~dex, and the RRab luminosity remains 
essentially unchanged ($\Delta\,M_V \simeq -0.001$~mag). 

For comparison purposes, we find, from the 
``bimodal'' HB simulations (Paper~I), the following mean values for all the 
RR Lyrae variables (RRab and RRc included): 
$\langle M_{\rm RR}\rangle = 0.618 \pm 0.002 \, M_{\odot}$, 
$\langle \log\,(L_{\rm RR}/L_{\odot})\rangle = 1.635 \pm 0.007$, 
$\langle \log\,T_{\rm e}^{\rm RR})\rangle = 3.829 \pm 0.006$ 
(the latter value is the average 
log-temperature; this corresponds to a $T_{\rm e}^{\rm RR} \simeq 6745$~K),  
$\langle M_V^{\rm RR}\rangle = 0.685 \pm 0.016$~mag.

\subsection{The Oosterhoff Type of M75}

As we have seen, the relatively large mean period of the M75 RRab, 
$\langle P_{\rm ab} \rangle = 0.5868$~d, and the high number  
fraction of RRc variables, $N_{\rm c}/N_{\rm RR} = 0.342$, do not clearly 
place M75 in the OoI group---as would be expected from its relatively high 
metallicity. In fact, as can be seen from Figure~7, the shift in periods 
with respect to M3, the prototypical OoI globular, seems to affect the 
whole distribution, not just the mean. 

The longer periods of the M75 RRab, compared to the other OoI globulars' 
mean RRab periods, could most plausibly 
be due to two effects: (i) A skewed temperature 
distribution, with the redder regions of the instability strip being 
preferentially populated in M75; (ii) A higher mean luminosity at any given  
temperature. The first possibility does not seem supported by our data, and 
would also appear inconsistent with the high fraction of RRc variables (which 
are, of course, bluer than the RRab). The second possibility, on the other 
hand, could explain the high $R$-ratio for the cluster, as found in Paper~I: 
For instance, an increase in the initial helium abundance for M75 would lead 
to both a higher $R$-ratio and to a brighter HB, thus implying longer periods 
at any given temperature. 

Clement \& Shelton (1999) have 
found that the Bailey $A_V - \log\,P$ diagram for stars satisfying the 
Jurcsik-Kov\'acs ``compatibility condition'' (i.e., with low values of 
$D_{\rm m}$) provides a useful means to obtain the Oosterhoff class of a 
GC. How does this diagram look in the case of M75, once the ``anomalous'' 
stars (i.e., those with values of $D_{\rm m} > 5.0$) have been removed? 

To answer this question, we had to pay 
special attention to the cases of V14, NV5, and NV9, for the $\chi$ values 
of these stars in the $V$ data were 4.1, 13.9, and 3.9, respectively, so 
they did not meet the criterion of $\chi < 3.5$ (\S2.2). The errors in their 
$V$ magnitudes were 0.0272, 0.1039, and 0.0801, respectively. We have used 
the errors in magnitude to recalculate the amplitudes assuming the stars 
in the {\sc isis} reference image were brighter or dimmer by the amount 
corresponding to the errors in magnitude. For completeness, error bars 
were similarly computed for NV2 and NV3. 

The resulting $A_V - \log\,P$ diagram is given in Figure~8. 
While the number of stars satisfying the 
compatibility condition is small, one can clearly see that the stars do 
not preferentially cluster around either of the OoI or OoII lines (kindly 
provived by C.~Clement). Therefore, if we rely on the Clement \& Shelton   
(1999) Oosterhoff classification scheme, we again find an indication that 
M75 is not of OoI type, better classifying as an Oo-intermediate globular. 

Borissova et al. (2001) found some surprising evidence that the position of 
a GC, in the Bailey diagram, may depend on whether $B$ or $V$ amplitudes are 
used. Is this the case for M75 as well? 

In Figure~9, we plot the M75 Bailey diagram, focusing again on the RRab 
stars only. We used a gray-tone scheme for the symbols, with variables 
having largest $D_{\rm m}$ values plotted in lighter gray. Unfortunately, 
the C.~Clement lines are not available in the $A_B - \log\,P$ plane, so 
that we must use different reference lines as representative of the two 
Oosterhoff classes. Therefore, we followed the same approach as used in 
producing Figure~5. 

Intriguingly, inspection of Figure~9 does not lead to a similar conclusion 
as Figure~8. The variables with ``regular'' light curves are now found to 
scatter around the OoI line, which is based on the M3 line derived by 
Borissova et al. (2001). NV3, which occupied an Oo-intermediate position 
in Figure~8, lies on a ``sub-OoI'' position in the new diagram. NV9, which 
was close to the OoII line, is now shifted down to a position close to the 
OoI line. The reason for this behavior is unclear, although it should be 
noted that the Borissova et al. M3 lines do differ from the C.~Clement OoI 
lines, even when the $V$ amplitude is used. This is likely due to the fact 
that Clement based her selection criteria on the Jurcsik-Kov\'acs 
compatibility criterion, whereas Borissova et al. only required that the 
variables showed no obvious indication of the Blazhko effect, or any other 
clear problems with their light curves. 

As far as the OoII line goes, we can only 
point out that the relation used to transform $V$ amplitudes into $B$ 
amplitudes, from Layden et al. (1999), is probably not valid for the 
M75 variables, predicting too large $B$ amplitudes for a given $V$ 
amplitude. The Layden et al. relation was based on the Dickens (1970) 
photographic data for NGC~6171 (M107) stars. Using the Layden et al. 
relation we find that the derived $V$ amplitudes of the M75 RRab variables 
with $D_{\rm m} < 5.0$ are underestimated, in the mean, by 
$\Delta\,A_V \approx 0.18 \pm 0.18$ (standard deviation). In any case, 
if our $V$ amplitudes are somehow incorrect due 
to problems with the derived {\sc isis} light curves, the corresponding 
$D_{\rm m}$ values, based as they are on the same $V$ light curves used 
to derive the $V$ amplitudes, should likely be very high---which they 
are not.   

In order to further investigate this issue, 
we selected the variable stars from Kaluzny et al. (1998)
with small $D_{\rm m}$ values, since those appear to have been the ones 
primarily used by Clement \& Shelton (1999) to derive their OoI line. These 
are shown in Figure~10a, in the $A_V - \log\,P$ plane. A few stars which, 
in spite of having small $D_{\rm m}$ values, are Blazhko variables 
according to Cacciari, Corwin, \& Carney (2003), are highlighted. 
The plot looks 
similar, though not identical, to Figure~1, top panel, in Clement \& 
Shelton. In general, there is little scatter around Clement's line, even 
if the Blazhko variables with small $D_{\rm m}$ are included. One can see 
the two stars that Clement \& Shelton call ``Oosterhoff II'' as well; they 
fall very close to Clement's ``OoII line,'' which is actually based on 
$\omega$~Cen stars.

Next we plotted the very same diagram, for the very same stars,
but using instead the $V$ amplitudes from Corwin \& Carney (2001). The 
result can be seen in Figure~10b. One now finds a lot
more scatter around Clement's OoI line; four stars, instead of
two, deviate toward the OoII region (but two of them are
Blazhko variables, in spite of having small $D_{\rm m}$). If one
ignores these four stars, Clement's OoI relation still provides
a reasonable description of the mean behavior of the ``OoI
variables,'' though one must now realize that there is somehow
more scatter around the line than suggested in the Clement \& Shelton 
(1999) original ``Oosterhoff-classification'' plot.

Finally, we again plotted the very same stars, but now in the $A_B - \log\,P$
diagram, in order to further investigate the reason for the differences 
between the M75 results in the diagrams using $A_V$ and $A_B$ (Figs.~8 and 9, 
respectively). The plot for M3 is shown in Figure~10c. The
$B$ amplitudes come from Corwin \& Carney (2001); Kaluzny et al. (1998) 
provided no $B$ amplitudes for their stars. For the ``OoI line,'' we used 
both the line that Cacciari et al. (2003) have obtained from the Corwin 
\& Carney data, and the line from Borissova et al. (2001); they are very 
similar. For the OoII line, we transformed Clement's
relation to $B$ amplitudes using the relation from Layden et al. (1999),
$A_V=0.72\,A_B+0.03$. What we find here may be relevant for the correct 
interpretation of the results from Figures~8 and 9. The main indications 
from Figure~10c are as follows: i)~There is significant scatter, and the 
clustering of stars around the ``Oo lines'' does not appear as clear-cut 
as in Clement \& Shelton (1999); ii)~In this plane, the OoI line tends to 
provide more of an ``upper bound'' to the stars used by Clement \& Shelton to 
obtain their OoI line than an actual description of their mean locus;
iii)~The ``OoII line'' based on the Layden et al. approach clearly fails 
to account for what Clement \& Shelton call ``OoII stars''
in M3; such a line seems again to provide just an upper bound
to the locus occupied by the ``OoII stars.'' A possible explanation 
may be that the Layden et al. relation may not be valid for M3 stars, 
either.

From these conclusions, and in order to obtain a completely
equivalent diagnostic of Oosterhoff type as originally provided
by Clement \& Shelton (1999), we should probably apply slight shifts 
to both the OoI and OoII lines in the $A_B - \log\,P$ diagram. If we 
do this, the ``Oo-intermediate'' nature of M75, as
suggested by the Clement classification scheme in the
$A_V - \log\,P$ diagram, would perhaps be more strongly supported 
by the $A_B - \log\,P$ diagram as well. The main problem with
this, of course, would be to interpret the reason why the OoI line, 
for the stars that Clement \& Shelton used, might have to be slightly 
shifted towards smaller amplitudes in the $A_B - \log\,P$ plane, 
compared to the lines provided by Cacciari et al. (2003) and by 
Borissova et al. (2001).\footnote{As pointed out by the referee, it is
important to note that the Oosterhoff classification according to the 
distribution of variable stars in the Bailey diagram is somewhat hampered 
by the scatter introduced by Blazhko RR Lyrae. The reason why the OoI line 
derived by Clement \& Shelton (1999) needs to be shifted toward smaller 
amplitudes might depend on this effect. As a matter of fact, RR Lyrae 
stars attain their largest amplitude in the $B$ band, and therefore the 
spread in amplitude due to Blazhko RR Lyrae is larger in the $B$ band 
than in any other band. This means that accurate estimates of Oo mean 
lines in the $A_B$ vs $\log\,P$ plane may require even more accurate 
evaluations of the RR Lyrae affected by Blazhko RR Lyrae. On the other 
hand, the $B$ amplitudes themselves are also greater, so that the 
fractional effect upon scatter in the Bailey diagram may be little 
different for $V$ and $B$.}

Therefore, the Bailey diagram neither strongly supports nor rules out an 
Oosterhoff-intermediate classification for M75. Do the derived physical 
parameters for the RR Lyrae variables, based on the Fourier decomposition 
of their light curves, suggest that they are intermediate between OoI and 
OoII? 

To answer this question, we compare the physical parameters of M75 with 
those similarly derived for other GCs in Tables~6 (RRc) and 7 (RRab). These 
tables represent up-to-date extensions of the compilations previously 
presented by Clement \& Rowe (2000) for the RRc stars and by Kaluzny et al. 
(2000) for the RRab stars. To the RRc entries from Clement \& Rowe, we have 
added values for NGC~6362 (Olech et al. 2001), M75 (this paper), NGC~6229 
(Borissova et al. 2001), NGC~6934 (Kaluzny, Olech, \& Stanek 2001), and 
NGC~2298 (Clement, Bezaire, \& Giguere 1995). To the RRab entries from 
Kaluzny et al., we have added values for NGC~6362, M75, NGC~6229, and 
NGC~6934. In both Tables~6 and 7, the 
clusters are listed in order of decreasing metallicity, the latter having 
been obtained from the Harris (1996) catalogue. The metal-rich GCs 
NGC~6388 and NGC~6441, whose Oosterhoff class is unclear (Pritzl et al. 
2001, 2002), are listed for completeness only; Trimble \& Aschwanden (2001) 
classified them as ``Osterhoff type III,'' expanding on the suggestion by 
Pritzl et al. (2000) that these may be prototypes of a new Oosterhoff 
class, whose mean RRab periods are even longer than for OoII globulars.  

Both Table~6 and Table~7 indicate normal values for the M75 RR Lyrae 
physical parameters, given its metallicity. Therefore, if one uses 
such parameters in order to provide guidance in obtaining the 
Oosterhoff classification of M75, the conclusion would be that the physical   
properties of the M75 variables are consistent with the cluster being OoI. 
In this case, the cluster's HB not being anomalously bright with respect to 
other OoI clusters, and the mean temperature of the RRab's also being similar 
to that for OoI clusters, we would be at a loss to explain why the RRab 
variables have longer periods than is typical for an OoI globular, and why 
its ``normal'' stars are shifted with respect to the OoI locus in the 
$A_V - \log\,P$ plane. 

If M75's HB is indeed not anomalously bright, the 
cause of the high $R$-ratio that we found in Paper~I 
must be related to the selective absence of bright red giants (compared to 
HB stars) in the regions over which the number counts were performed. 
On the basis of the current photometry, it seems that RGB stars could 
have a different radial 
distribution with respect to HB stars. This effect could be due
to an observational bias or to the internal dynamics of the cluster.
In order to clarify this issue, we are currently using HST archive data. 
The results of such an analysis, along with a full reevaluation of the 
number counts reported in Paper~I, will be given in Ferraro et al. (2003, 
in preparation---Paper~III). In any case, our preliminary 
revision of the HB number ratios indicate that the HB 
bimodality may be even more pronounced than indicated in Paper~I. 

Furthermore, if the brightness of M75's HB is comparable to that in other 
OoI globulars, it immediately follows, from the discussion in Catelan,  
Sweigart, \& Borissova (1998), that whatever the second parameter(s) 
that cause the HB 
bimodality in M75, it (they) must not have changed the brightness of the 
HB stars. Most second parameter candidates move stars vertically in the 
CMD at the same time as they ``slide'' them horizontally along the H-R 
diagram; this is the case, in particular, with the original helium 
abundance, helium mixing and core-mass changes (as due, e.g., to rotation 
of the cores of RGB stars; Mengel \& Gross 1976). If any of these second  
parameter candidates were at play, being responsible for the extension 
of M75's HB farther to the blue than is commonly the case among GCs of 
this metallicity, we would expect M75's RR Lyrae to be brighter than in 
other OoI globulars. While this would be consistent with an 
Oo-intermediate classification for this cluster, it would obviously not 
explain why the Fourier decomposition parameters do not indicate the M75 
RR Lyrae to be anomalously bright. In order to achieve a fully consistent 
picture, one might be forced to conclude that there are problems with the 
methods used to derive the physical parameters from the Fourier decomposition 
of the light curves, both for RRc and RRab variables. In fact, this is not 
totally unlikely: Challenging problems are indeed known to exist at the 
RRab side (e.g., Koll\'ath, Buchler, \& Feuchtinger 2000), whereas, at the 
RRc camp, it is unclear whether the models used by Simon \& Clement (1993) 
yield results that are consistent with the current generation of 
hydrodynamical RR Lyrae models; in fact, there is evidence that the  
Simon \& Clement relations already break down for the RRc star U~Comae 
(Bono, Castellani, \& Marconi 2000). 
Unfortunately, it appears that one will have to wait until the Fourier 
decomposition methods of derivation of the physical parameters of RR Lyrae 
stars are placed on a more solid footing, before conclusively determining 
the Oosterhoff type of M75.

\section{Summary}
In the present paper, we have presented the first extensive CCD investigation of 
the variable star population in the GC M75 (NGC~6864). Several stars previously 
listed as variable and suspected to be RR Lyraes (V5, V6, V7, V10, and V15), do 
not appear to be variable on the time scale of a few days. We were able to 
derive periods and $B$, $V$ light curves for nine priviously known RR Lyrae 
variables and for 29 newly discovered RR Lyrae variables. About half of the new 
discoveries we owe to an application of the image subtraction method developed 
by Alard (2000) and Alard \& Lupton (1998), which does appear to be a very 
powerful tool to search for new variables and derermine their periods, 
especially in crowded regions such as the cores of GCs.

The confirmed RR Lyrae population in M75 consists of 13 RRc and 25 RRab stars. 
Although M75's metallicity is typical for an OoI cluster, its unusually high 
average RRab period of 0.5868~d, its ratio of RRc to RRab stars (which is 
intermediate between OoI and OoII clusters) and its unusual HB 
morphology distinguish it from other OoI clusters. A Bailey diagram for the 
cluster is plotted in Figure~5 and variables whose position in the diagram 
appear to be anomolous are discussed. The positions of the (few) M75 variables 
which satisfy the compatibility criterion of Jurcsik \& Kov\'acs in the Bailey 
diagram also suggests that the cluster may be of Oo-intermediate type, since 
these variables do not cluster around the areas occupied by most RRab stars in 
either OoI or OoII clusters. However, the mean luminosity of the variables, 
according to the Fourier decomposition of their light curves, does not indicate 
a deviation from other ``normal'' OoI globulars. If confirmed, this would also 
imply that the anomalous $R$-ratio of M75, as measured in Paper~I, cannot be 
due to a high helium abundance; dynamical effects that somehow affect only 
RGB mass loss would have to be invoked. A more detailed investigation of the 
radial gradients and other indicators of dynamical effects in M75 will be 
carried out in the next paper of this series (Paper~III).

\acknowledgments Christine Clement is warmly thanked for providing us with 
her Oosterhoff lines in the $A_V - \log\,P$ diagram. We thank the referee, 
Giuseppe Bono, for his useful suggestions and perceptive comments. Support 
for M.C. was provided by Proyecto de Inicio DIPUC 2002-04E. H.A.S. thanks the 
US National Science Foundation for support under grant AST99-86943. J.B. is 
supported by FONDAP Center for Astrophysics grant number 15010003. F.R.F. 
acknowledges the financial support of the {\it Agenzia Spaziale Italiana} 
(ASI) and  the {\it Ministero della Ricerca Scientifica e Tecnologica} (MURST).

\appendix
\section{Notes On Individual Stars}

V2, PRC82 -- This star is bright and red relative to the RR Lyraes. It is not 
variable on the scale of a few days.

V5, PRC82 -- This star is near the horizontal branch (HB) and red relative to the 
RR Lyraes. It is not variable on the scale of a few days.

V6, PRC82 -- This star is near the HB and red relative to the RR Lyraes. It is 
not variable on the scale of a few days.

V7, PRC82 -- This is two bright stars 3.6\arcsec apart. Neither star is variable 
on the scale of a few days.

V10, PRC82 -- This star is near the HB and blue relative to the RR Lyraes. It is 
not variable on the scale of a few days.

V15, PRC82 -- This star is near the HB and red relative to the RR Lyraes. It is 
not variable on the scale of a few days.

V16 and V17, PRC82 -- These stars are red giants.

S1, S2, S3, and S4, PRC82 –- These stars were checked for variability and found
to be not variable on the scale of a few days. 

NV3 -- This RRab variable lies well into the RRc region of the instability strip 
(Fig.~6). This, together with the fact that its brightness places it above the mean  
level and that it has an unusually low ratio of the $B$ to $V$ amplitudes may 
indicate that it is a blended star. 

NV7 -- This RRab variable has a color well to the red of the instability strip 
(Fig.~6). This, together with the fact that its brightness places it above the mean 
level and that it has a high ratio of the $B$ to $V$ amplitudes may indicate that it 
is a blended star. Its $D_{\rm m}$ value is high.  

NV8 -- This RRc variable is somewhat anomalous in Figure~4, in that it has a 
relatively long period for an RRc variable (0.3776 day) and yet it has a 
relatively high surface temperature ($\bv = 0.236$ mag). It also has an average 
$V$ magnitude well above the ZAHB level and is probably blended with a bluer star.
The error in the phase difference $\phi_{31}$ does not stand out in comparison 
with the other RRc stars. 

NV10 -- This RRab variable has a unique position in the period-amplitude plane 
(Fig.~4). Its period is larger than for other variables with comparable 
amplitudes, indicating a more luminous star. It is in fact somewhat brighter 
than the mean for the other stars, as can be seen from Figure~6. 
This might indicate that it is a more highly evolved star, in 
which case it might be the progeny of one of the lower-mass modes of the M75 
ZAHB mass distribution (see Paper~I). NV10 has an appropriate position on the 
color-period plot (Fig.~3). Fourier analysis of its light curve was hampered 
by the presence of a gap in phase coverage in the range 
$0.25 \lesssim \phi \lesssim 0.45$, which led to a spurious fit in that region. 

NV17 -- This RRab variable is at a position in the period-amplitude plane (Figure 
4) that indicates an underluminous star. However its intensity-averaged $B$ 
magnitude indicates that it is bright relative to the other RR Lyraes. This 
might be explained by the fact that it is a foreground star, a blended star, or 
that it is a Blazhko variable at less than maximum amplitude. Its light curve 
suggests the possibility that it is a Blazhko variable. It has a value of 
$D_{\rm m} = 6.34$, which is one of the lowest among the RRab with ``anomalous''
light curves, according to the Jurcsik-Kov\'acs criterion. 

NV18 -- This variable has very unreliable photometry and its period and even its 
identification as an RRab variable are uncertain.

NV20 -- This RRab variable has unreliable photometry and its period is uncertain.

NV22 -- This RRab variable has unreliable photometry and its period is uncertain.

NV23 and NV24 -- Like NV17, these RRab variables are in positions in the 
period-amplitude plane (Figure 5) that indicate underluminous stars while 
they are in 
fact relatively bright. Possible explanations, as with NV17, include their being 
foreground stars, blended stars, or Blazhko variables at less than maximum 
amplitude. The (uncertain) Fourier parameters indicate that both NV23 and NV24
have ``anomalous'' light curves (i.e., large $D_{\rm m}$ values). 

NV26 -- This RRc variable has unreliable photometry and its period is uncertain.
 
NV28 -- This RRc variable has an extremely small amplitude. Its positions in 
Figures 2, 4 and 5 do not indicate anything unusual about it otherwise. Because 
of the large scatter in the data, its period is uncertain.

NV29 -- This RRc variable has very unreliable photometry and 
its period is uncertain. We have no $V$ data for this star.

NV 30 - This variable has very unreliable photometry and its period and 
variability type are uncertain. The shape of its light curve resembles 
that of an RS~CVn variable. We have no $V$ data for this star.

\clearpage

\begin{figure*}[t]
  \figurenum{1a}
  \centerline{\psfig{figure=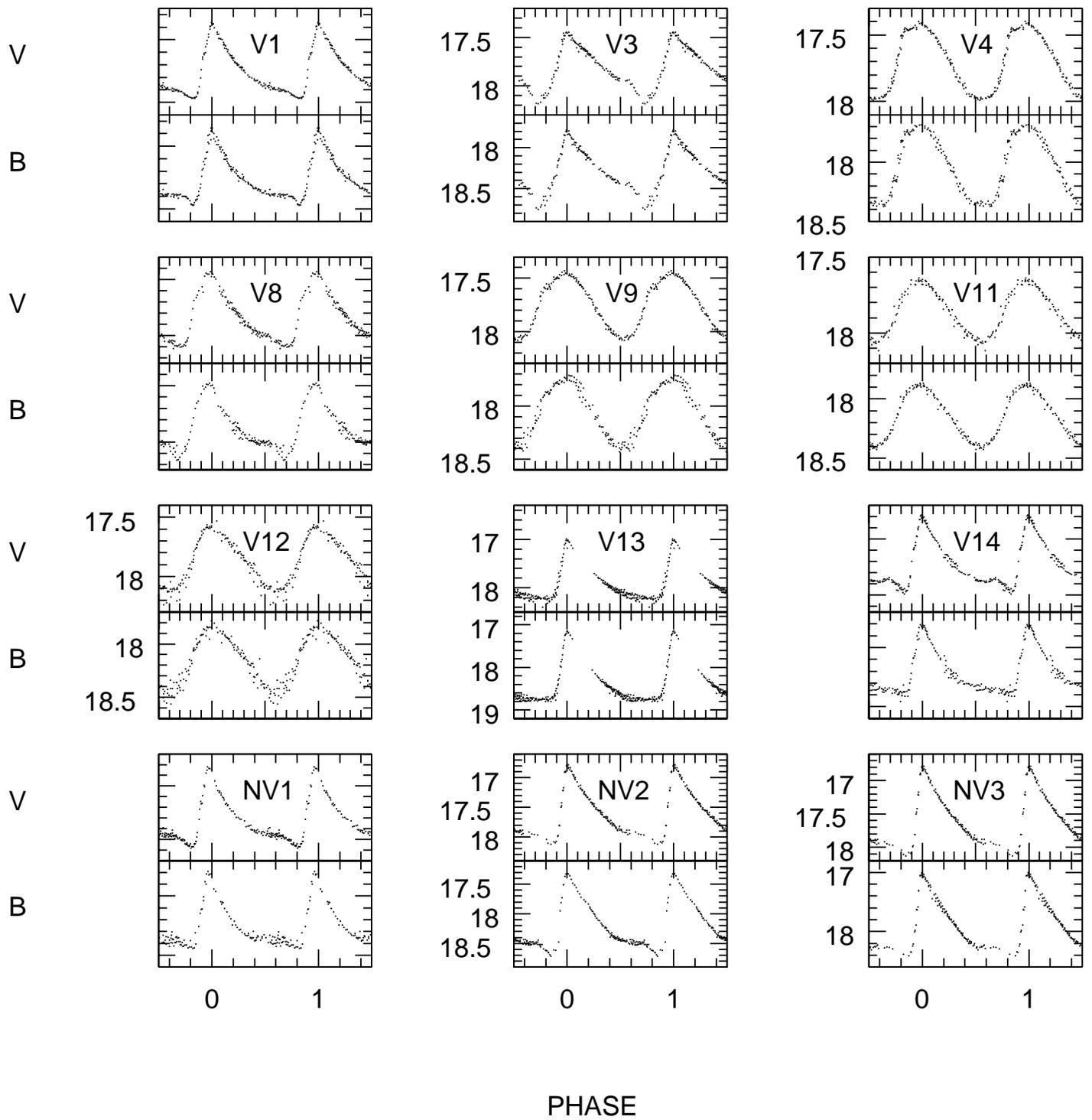}}
  \caption{Light curves for the variable stars in M75.
      }
      \label{Fig01a}
\end{figure*}

\begin{figure*}[t]
  \figurenum{1b}
  \centerline{\psfig{figure=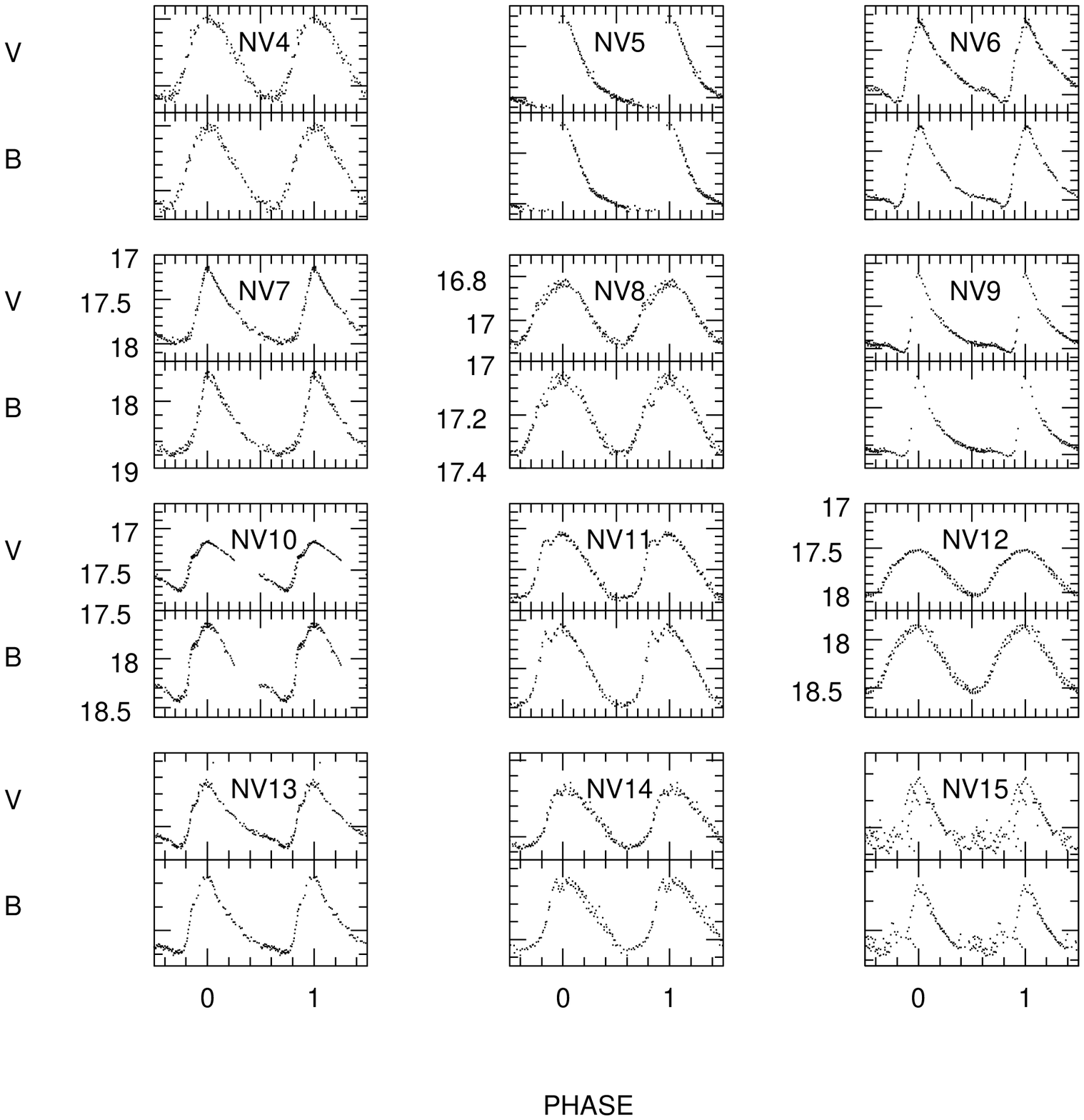}}
  \caption{Light curves for the variable stars in M75 (cont.). 
      }
      \label{Fig01b}
\end{figure*}

\begin{figure*}[t]
  \figurenum{1c}
  \centerline{\psfig{figure=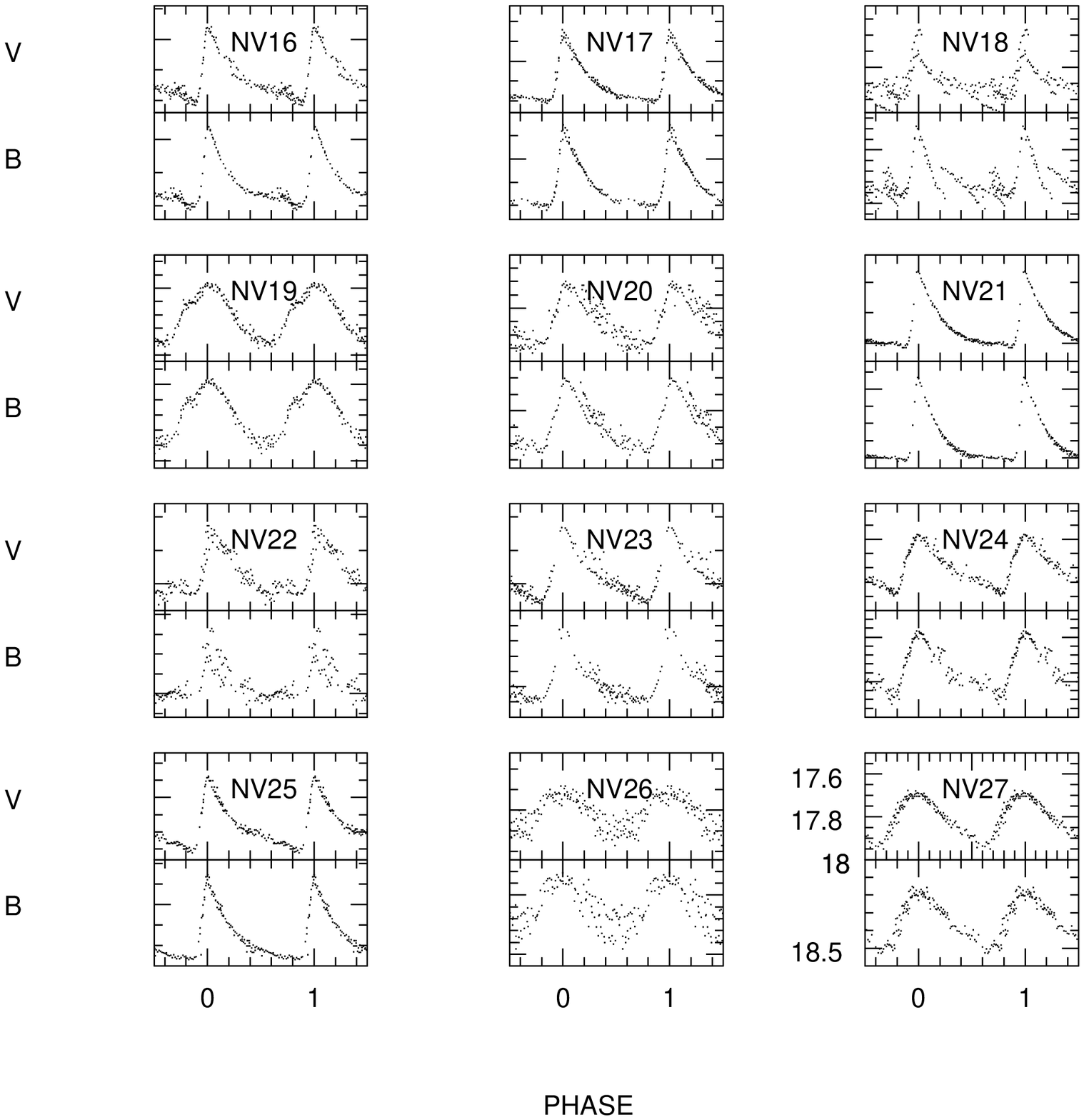}}
  \caption{Light curves for the variable stars in M75 (cont.).
      }
      \label{Fig01c}
\end{figure*}

\begin{figure*}[t]
  \figurenum{1d}
  \centerline{\psfig{figure=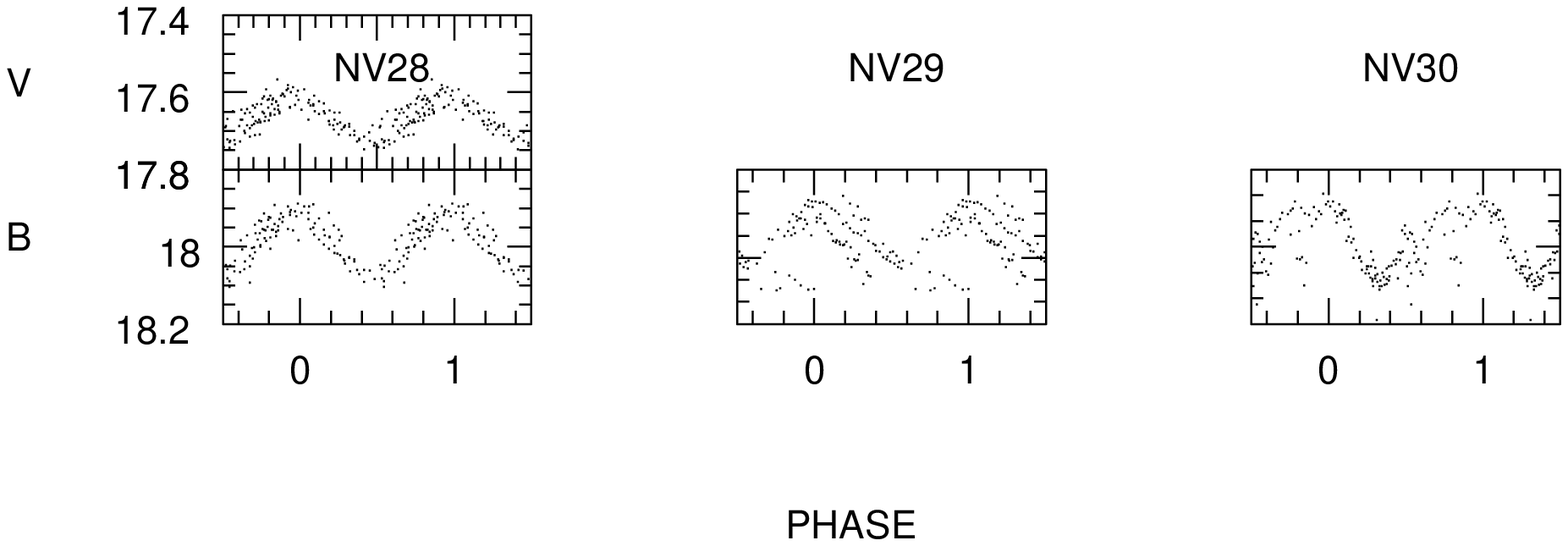}}
  \caption{Light curves for the variable stars in M75 (cont.).
      }
      \label{Fig01d}
\end{figure*}

\clearpage

\begin{figure*}[t]
  \figurenum{2}
  \centerline{\psfig{figure=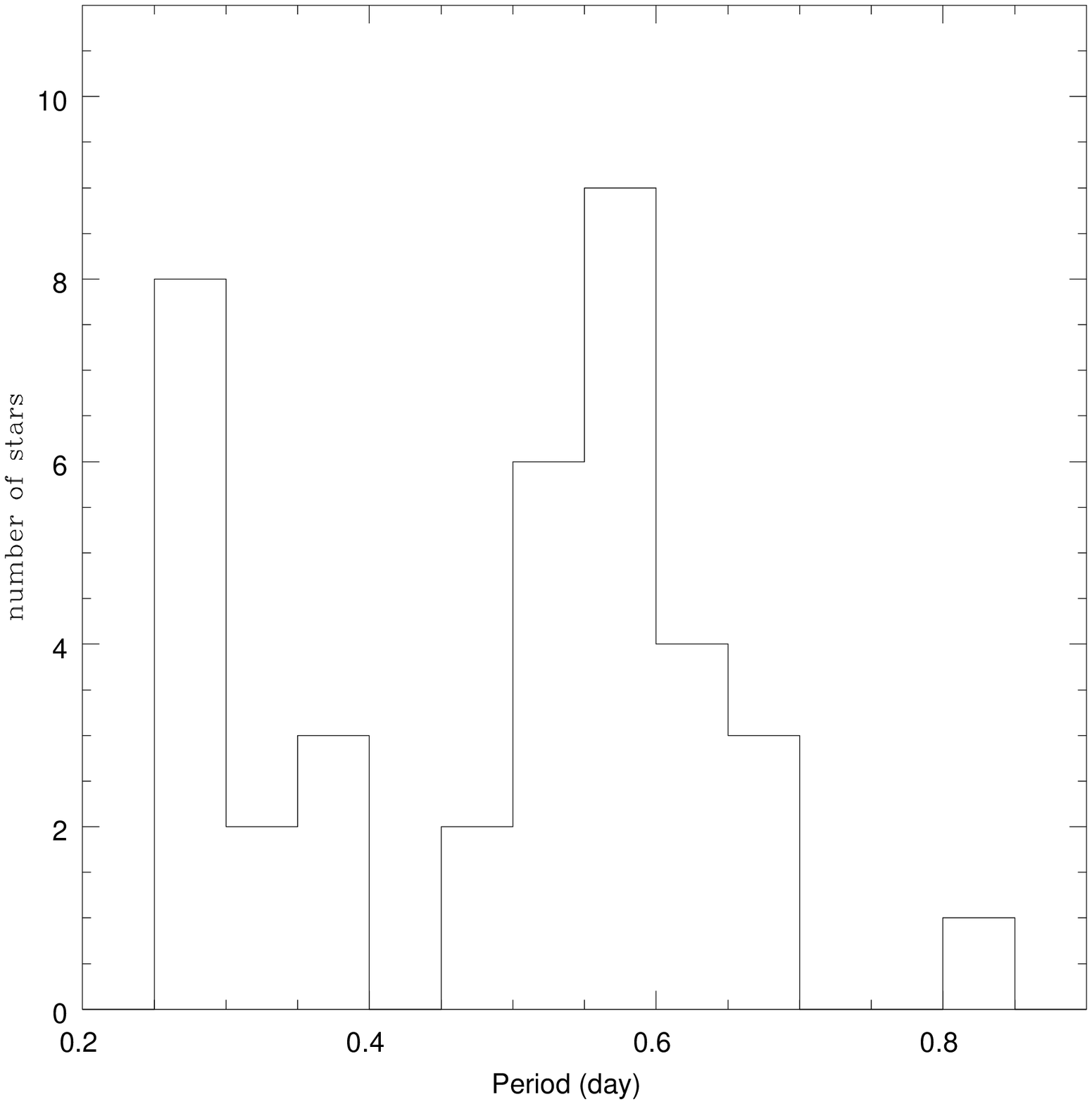,height=15cm,width=15cm}}
  \caption{Histogram of the periods of M75 variables.
      }
      \label{Fig02}
\end{figure*}

\clearpage

\begin{figure*}[t]
  \figurenum{3}
  \centerline{\psfig{figure=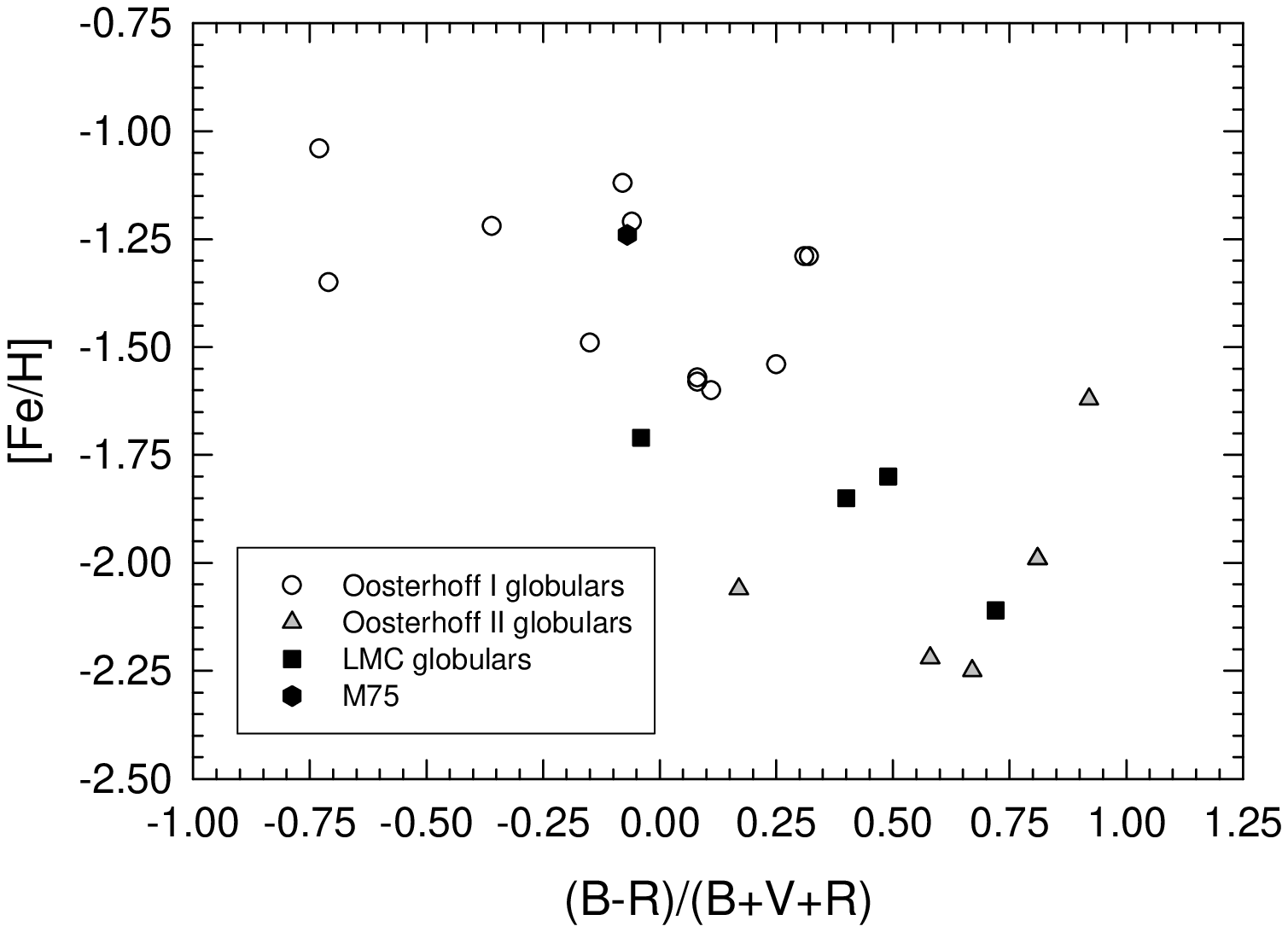}}
  \caption{HB morphology-metallicity plane, with the position of 
     M75 (filled hexagon) and LMC globulars (filled squares) 
     shown alongside OoI (open circles) and OoII (gray triangle) 
     GCs. Note that M75 is much more metal-rich than the 
     Oo-intermediate LMC globulars. 
      }
      \label{Fig03}
\end{figure*}

\clearpage

\begin{figure*}[t]
  \figurenum{4}
  \centerline{\psfig{figure=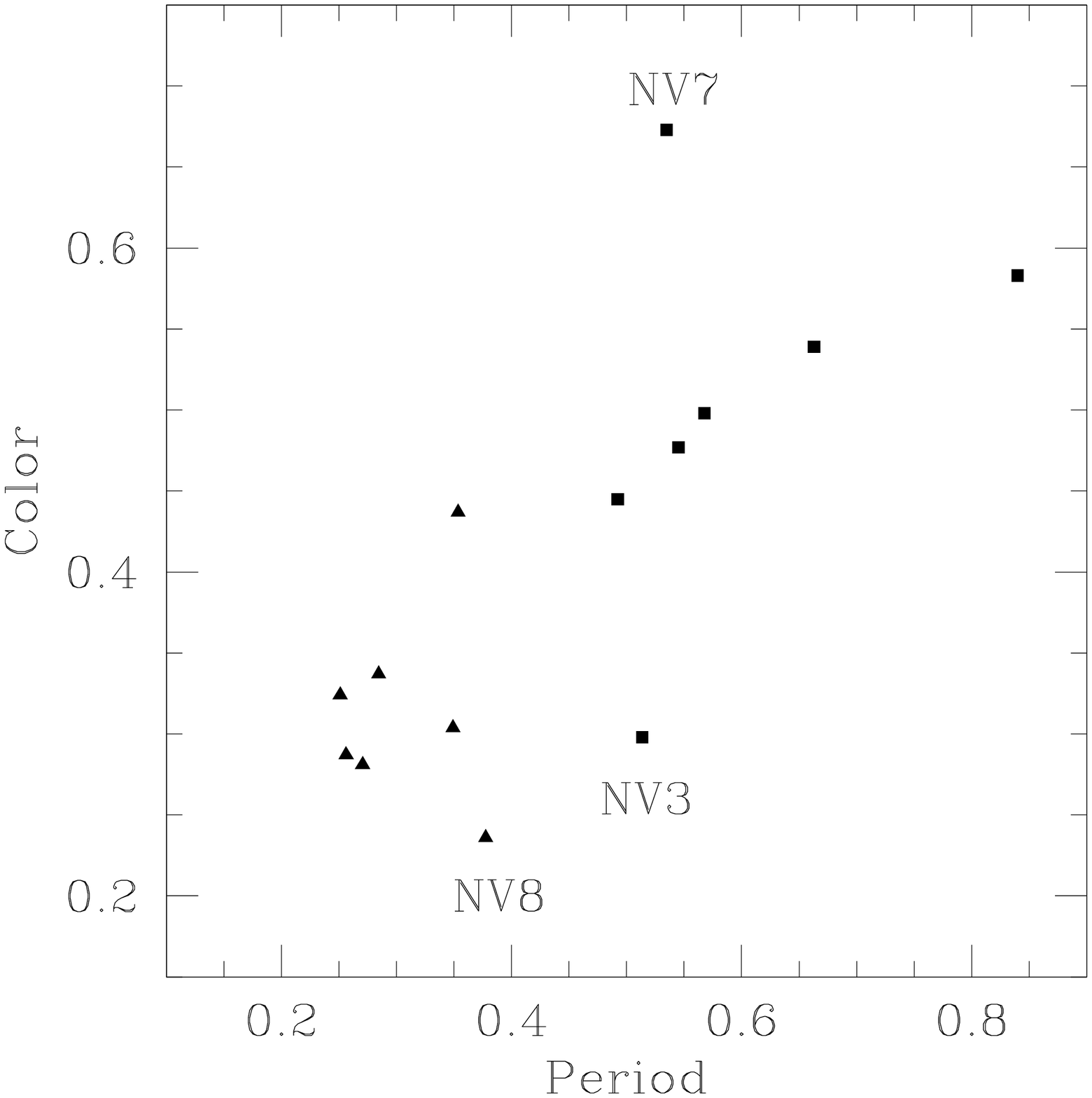,height=15cm,width=15cm}}
  \caption{Color vs. period diagram. Filled triangles indicate the
     RRc's, whereas filled squares are used for the RRab's.
     }
      \label{Fig04}
\end{figure*}

\clearpage

\begin{figure*}[t]
  \figurenum{5}
  \centerline{\psfig{figure=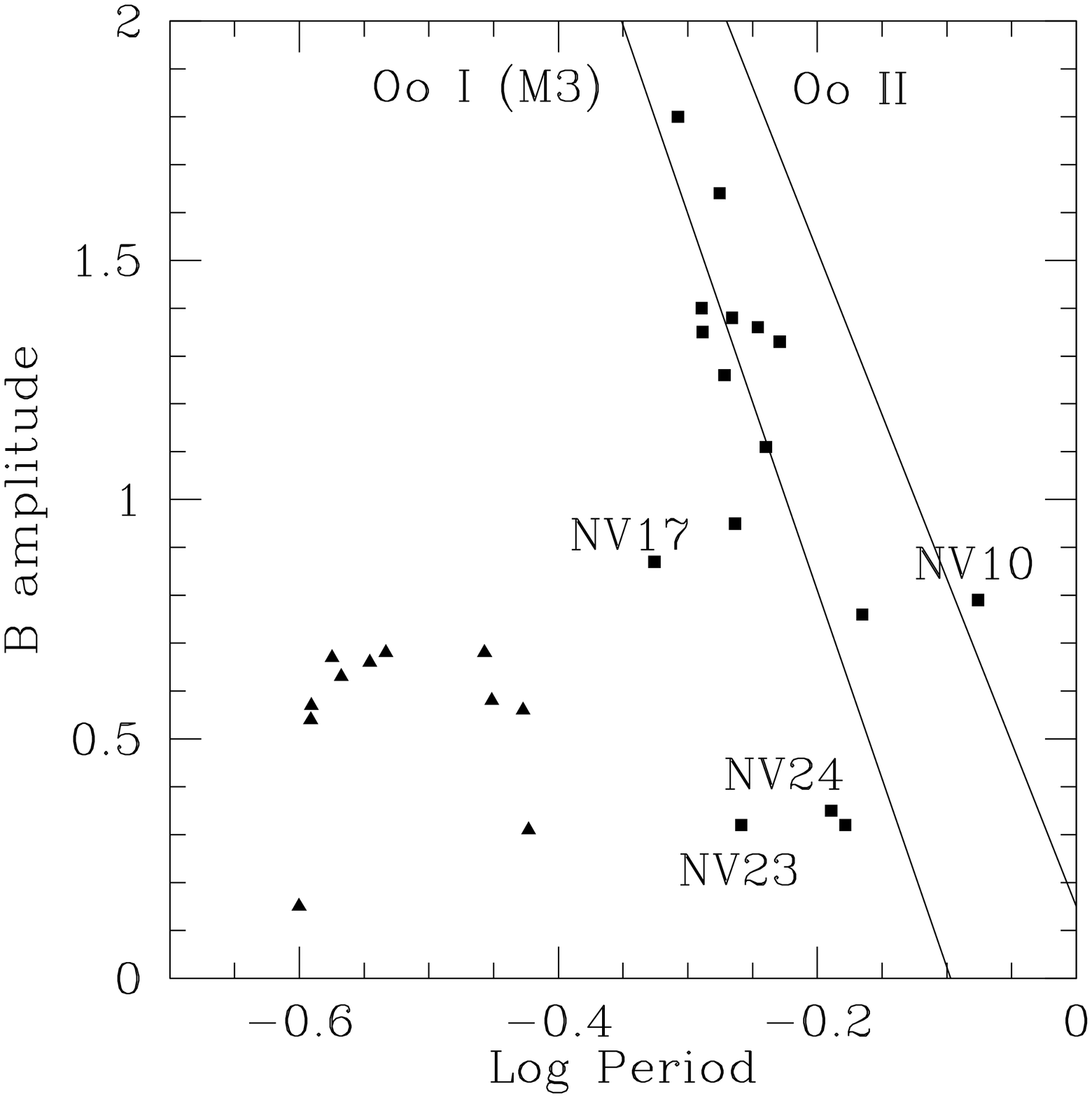,height=15cm,width=15cm}}
  \caption{$B$ amplitude vs. log period diagram (symbols as in 
     Figure~4). 
      }
      \label{Fig05}
\end{figure*}

\clearpage

\begin{figure*}[t]
  \figurenum{6}
  \centerline{\psfig{figure=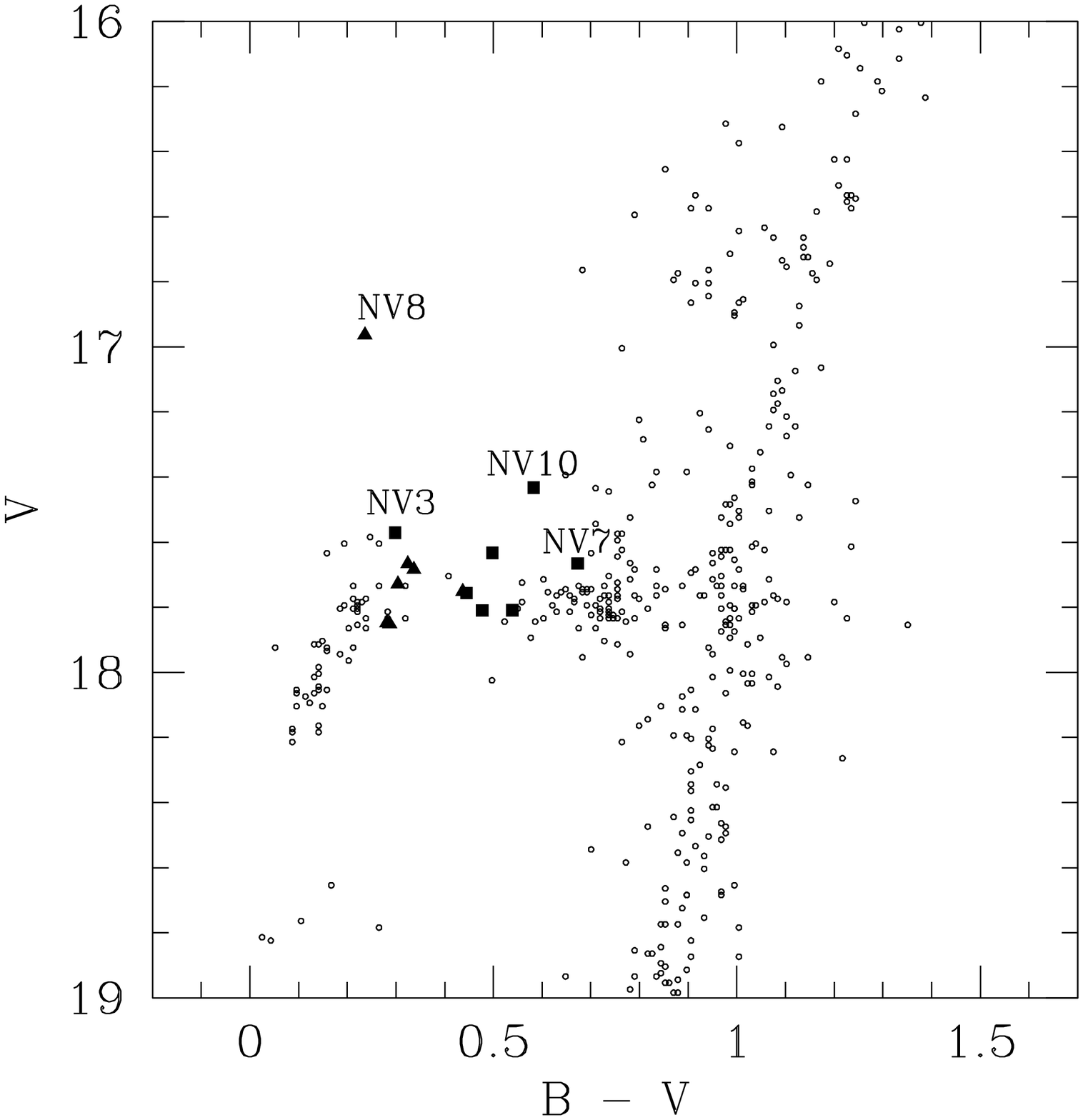,height=15cm,width=15cm}}
  \caption{Color-magnitude diagram for M75 with variables represented 
     by their intensity-weighted mean $V$ magnitudes and 
     magnitude-weighted mean (\bv) color indices (symbols as in 
     Fig.~4). 
      }
      \label{Fig06}
\end{figure*}

\clearpage

\begin{figure*}[t]
  \figurenum{7}
  \centerline{\epsfig{file=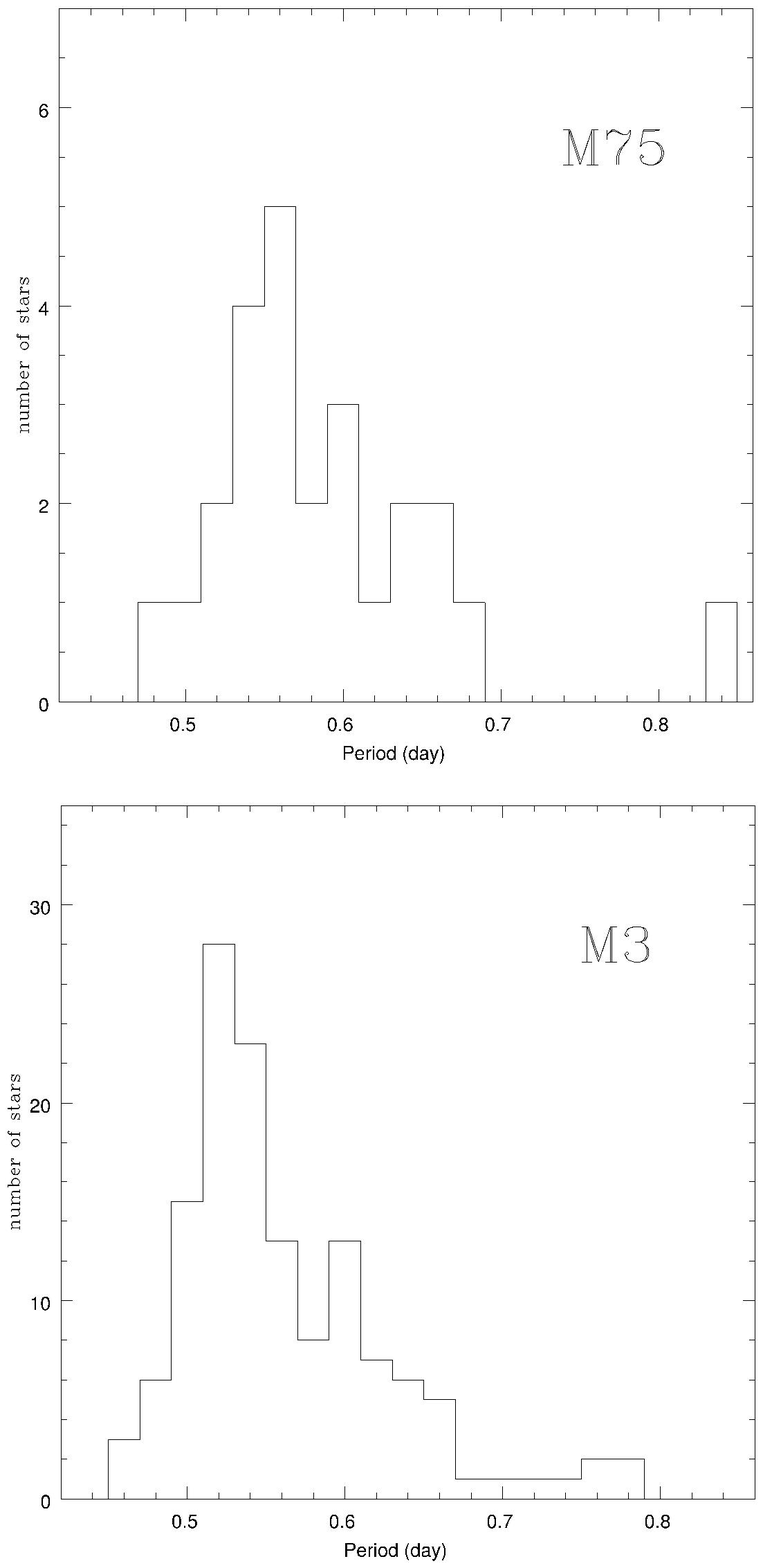,height=8in,width=4.25in}}
  \caption{Histograms for the RR Lyrae variables in M75 (upper panel) 
    and M3 (lower panel). RRc periods have been fundamentalized. 
    There is an indication that the whole of 
    the M75 period distribution is shifted towards slightly longer 
    periods. 
      }
      \label{Fig07}
\end{figure*}

\clearpage
\begin{figure*}[t]
  \figurenum{8}
  \centerline{\psfig{figure=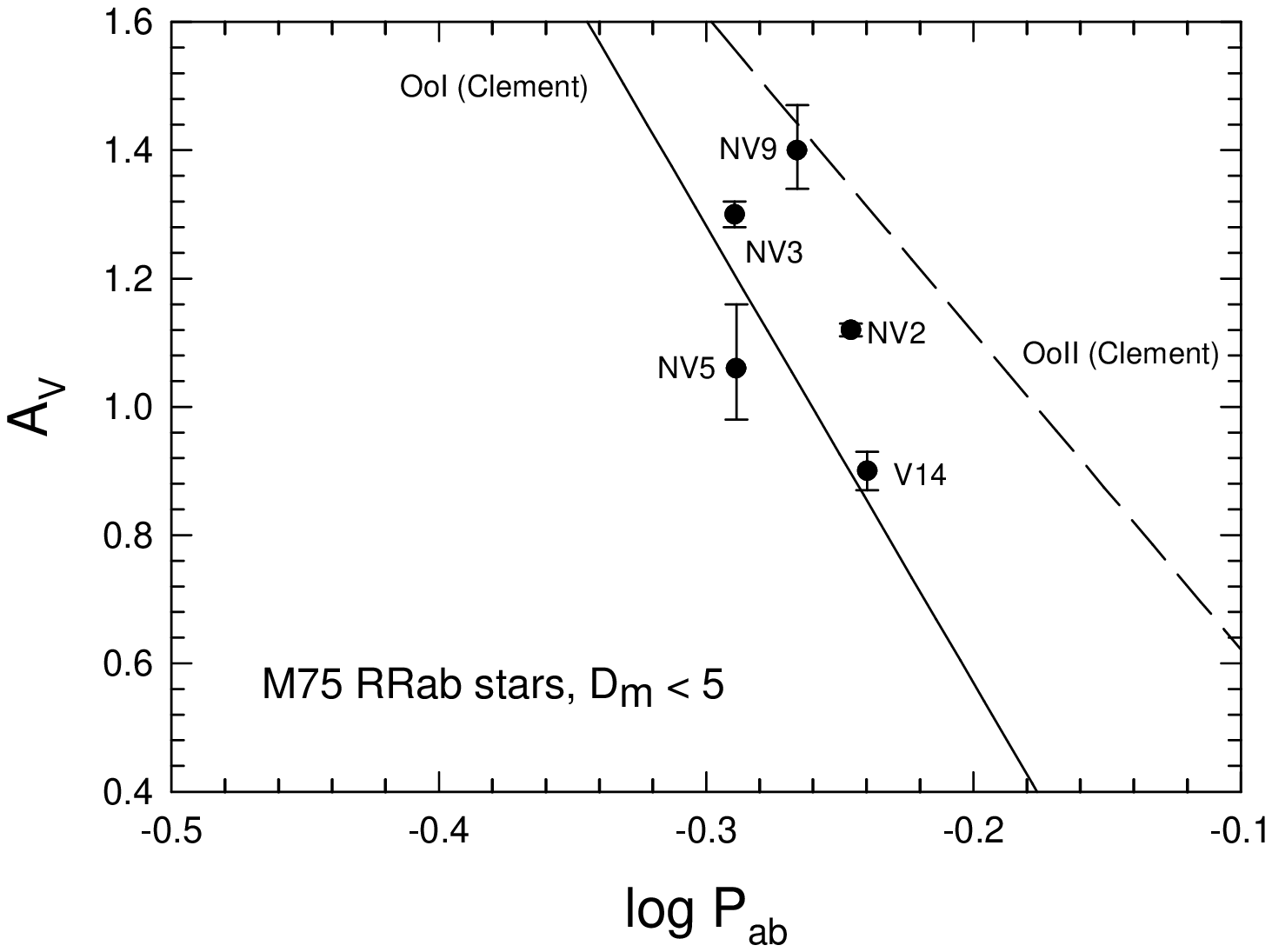}}
  \caption{Bailey diagram for M75 RRab stars, in the $A_V - \log\,P$ 
     plane. Only the variables with $D_{\rm m} < 5$ are shown. The OoI and 
     OoII ines are from Clement, and are similarly based on stars 
     with small $D_{\rm m}$ values. 
      }
      \label{Fig08}
\end{figure*}

\clearpage

\begin{figure*}[t]
  \figurenum{9}
  \centerline{\psfig{figure=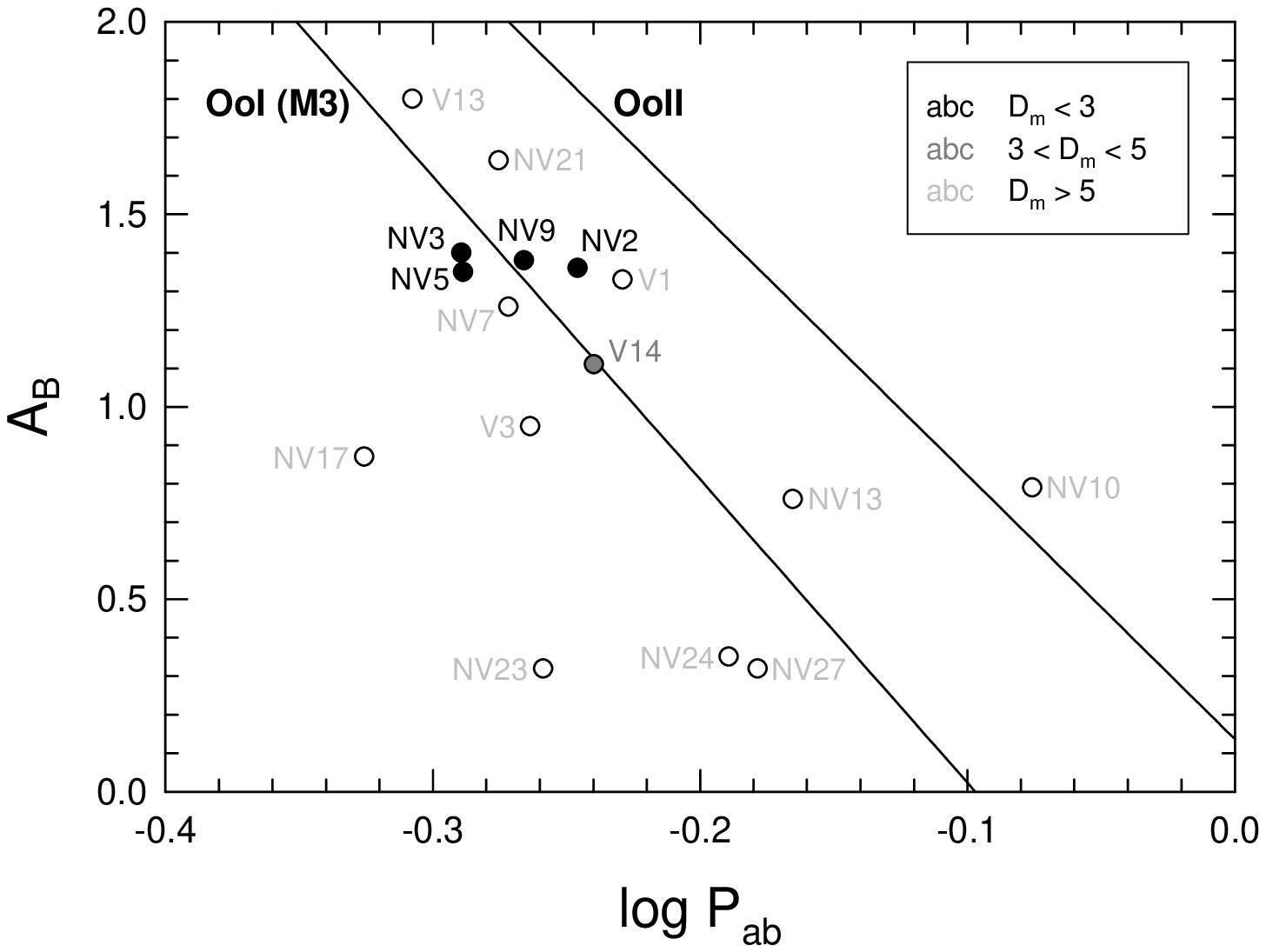}}
  \caption{Bailey diagram for M75 RRab stars, in the $A_B - \log\,P$ 
     plane. Variables with $D_{\rm m} > 5$ are shown as open circles; those 
     with $D_{\rm m} < 3$ as filled circles; and the one variable with 
     $3 < D_{\rm m} < 5$ as a gray circle. The ``OoI line'' actually 
     corresponds to the average M3 line, as obtained by Borissova 
     et al. (2001). 
      }
      \label{Fig09}
\end{figure*}

\clearpage

\begin{figure*}[t]
  \figurenum{10}
  \centerline{\epsfig{file=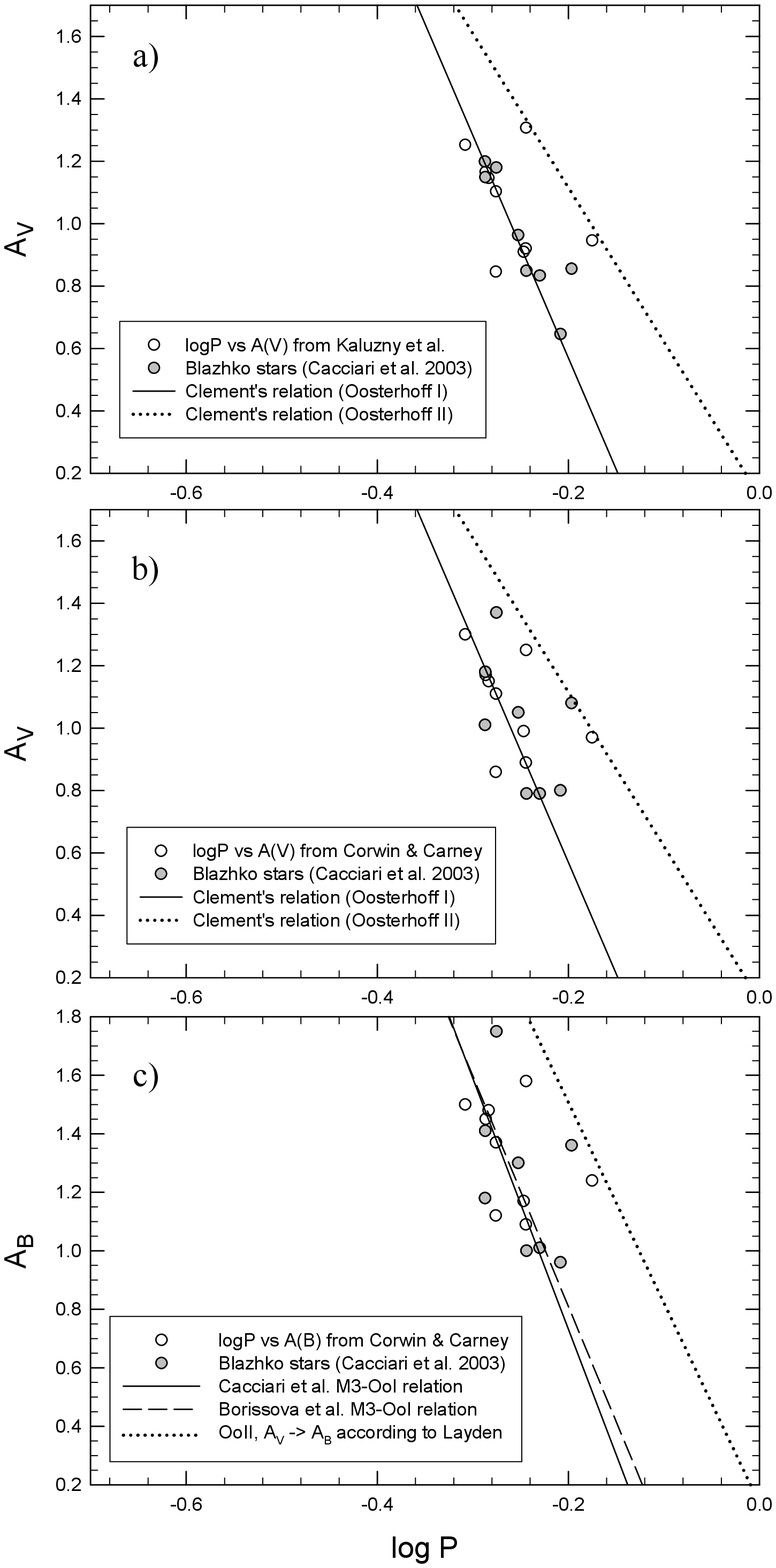,height=8.25in,width=4.25in}}
  \caption{Bailey diagram for M3 RRab stars, in the $A_V - \log\,P$
     (panels a and b) and $A_B - \log\,P$ (panel c) planes.
     Blazhko variables, as classified by Cacciari et al. (2003), 
     are indicated by gray circles. The meaning of the several 
     plotted lines is discussed in the text. 
      }
      \label{Fig10}
\end{figure*}

\clearpage

\begin{deluxetable}{lrrcccccccccl} 
\tablenum{1}
\tablewidth{0pc}
\footnotesize
\tablecaption{Ephemerides and Photometric Parameters for M75 Variables}
\tablehead{
  \colhead{ID} & 
  \colhead{$X(\arcsec)$} & 
  \colhead{$Y(\arcsec)$} & 
  \colhead{$P$(d)} & 
  \colhead{Epoch (JD)} & 
  \colhead{$A_{B}$} & 
  \colhead{$A_{V}$} & 
  \colhead{$\langle B_{\rm mag}\rangle$} & 
  \colhead{$\langle B_{\rm int}\rangle$} & 
  \colhead{$\langle V_{\rm mag}\rangle$} & 
  \colhead{$\langle V_{\rm int}\rangle$} & 
  \colhead{$(\bv)_{\rm mag}$} & 
  \colhead{Comments} 
}
\startdata
v1 &	14 &	-83 &	0.5901 &	2451379.696 &	1.33 &	\nodata &	18.311 &	18.250 &\nodata	  &\nodata  &\nodata 	&	RRab \\
v3 &	18 &	86 &	0.5451 &	2451379.501 &	0.95 &	0.73 &	18.308 &	18.278 &	17.831 &	17.810 &	0.477 &	RRab \\
v4 &	-18 &	-85 &	0.2847 &	2451379.904 &	0.66 &	0.57 &	18.040 &	18.011 &	17.703 &	17.683 &	0.337 &	RRc \\
v8 &	-29 &	-53 &	0.6503 &	2451379.403 &	\nodata &0.54 &	 \nodata&	\nodata &	17.812 &	17.797 & \nodata&	RRab \\
v9 &	43 &	-25 &	0.3492 &	2451379.713 &	0.70 &	0.63 &	18.055 &	18.030 &	17.751 &	17.729 &	0.304 &	RRc \\
v11 &	-121 &	85 &	0.2563 &	2451379.549 &	0.54 &	0.41 &	18.148 &	18.132 &	17.861 &	17.851 &	0.287 &	RRc \\
v12 &	37 &     74 &	0.2706 &	2451379.530 &	0.66 &	0.54 &	18.146 &	18.125 &	17.865 &	17.848 &	0.281 &	RRc \\
v13 &	127 &	-59 &	0.4924 &	2451379.867 &	1.65 &	1.30 &	18.285 &	18.136 &	17.840 &	17.756 &	0.445 &	RRab \\
v14 &	32 &	-6 &	0.5758 &	2451379.670 &	1.11 &\nodata &	18.223 &	18.163 &	\nodata	 & \nodata  &\nodata 	 &	RRab \\
				 							 	  
nv1 &	-9 &	56 &	0.5656 &	2451379.769 &\nodata &\nodata &\nodata 	 &\nodata 	 &\nodata  &\nodata 	 &\nodata	 &	RRab \\
nv2 &	1 &	39 &	0.5677 &	2451379.871 &	1.36 &	1.14 &	18.181 &	18.098 &	17.683 &	17.633 &	0.498 &	RRab \\
nv3 &	-7 &	25 &	0.5136 &	2451379.576 &	1.40 &	1.34 &	17.944 &	17.854 &	17.646 &	17.571 &	0.298 &	RRab \\
nv4 &	-27 &	8 &	0.2566 &	2451379.641 &	0.57 &\nodata &	17.960 &	17.943 &\nodata	 & \nodata	  &\nodata	 &	RRc \\
nv5 &	10 &	-17 &	0.5144 &	2451379.482 &	1.35 &\nodata  &18.110 &	17.999 &\nodata	  &\nodata	  &\nodata	 &	RRab \\
nv6 &	15 &	-20 &	0.6122 &	2451379.938 &\nodata &	0.73 &\nodata 	 & 	\nodata &	17.763 &        17.742 &\nodata	 &	RRab \\
nv7 &	-17 &	-24 &	0.5349 &	2451379.665 &	1.25 &	0.86 &	18.371 &	18.298 &	17.698 &	17.665 &	0.673 &	RRab \\
nv8 &	5 &	-25 &	0.3776 &	2451379.967 &	0.30 &	0.30 &	17.205 &	17.200 &	16.969 &	16.964 &	0.236 &	RRc \\
nv9 &	26 &	-36 &	0.5421 &	2451379.813 &	1.38 &\nodata &	18.096 &	18.014 &\nodata	  &\nodata 	  &\nodata	 &	RRab \\
nv10 &	-14 &	-40 &	0.8398 &	2451379.967 &	0.79 &	0.60 &	18.031 &	17.986 &	17.448 &	17.432 &	0.583 &	RRab \\
nv11 &	-34 &	-42 &	0.2930 &	2451379.663 &	0.68 &\nodata &	17.967 &	17.936 &\nodata	  &\nodata 	  &\nodata	 &	RRc \\
nv12 &	46 &	18 &	0.3536 &	2451379.765 &	0.68 &	0.50 &	18.204 &	18.178 &	17.767 &	17.752 &	0.437 &	RRc \\
nv13 &	19 &	13 &	0.6834 &	2451379.862 &	0.76 &\nodata &	18.319 &	18.290 &\nodata	  &\nodata 	  &\nodata	 &	RRab \\
nv14 &	15 &	4 &	0.2662 &	2451379.813 &	0.67 &\nodata &	17.971 &	17.943 &\nodata	  &\nodata 	  &\nodata	 &	RRc \\
nv15 &	6 &	-4 &	0.6262 &	2451379.600 &	\nodata &\nodata  &\nodata	 &\nodata 	 &\nodata  &\nodata 	 &\nodata &	RRab \\
nv16 &	0 &	-6 &	0.5651 &	2451379.720 &\nodata &\nodata  &\nodata	 &\nodata	 &\nodata  &\nodata	 &\nodata	 &	RRab \\
nv17 &	-10 &	-28 &	0.4724 &	2451379.680 &	0.87 &\nodata  &17.548 &	17.515	  &\nodata  &\nodata 	 &\nodata	 &	RRab \\
nv18 &	-11 &	5 &	0.5673 &	2451379.704 &	\nodata &\nodata  &\nodata	 &\nodata	 &\nodata  &\nodata	 &\nodata &	RRab \\
nv19 &	-13 &	-4 &	0.3739 &	2451379.761 &	0.56 &\nodata  &	17.659 &	17.640	  &\nodata  &\nodata 	 &\nodata &	RRc \\
nv20 &	-17 &	0 &	0.5969 &	2451379.643 &\nodata &\nodata  &\nodata	 &\nodata	 &\nodata  &\nodata	 &\nodata	 &	RRab \\
nv21 &	-17 &	-3 &	0.5304 &	2451379.908 &	1.64 &\nodata  &18.118 &	17.969	 &\nodata &\nodata 	 &\nodata	 &	RRab \\
nv22 &	-1 &	6 &	0.6273 &	2451379.978 &\nodata &\nodata  &\nodata	 &\nodata	 &\nodata  &\nodata	 &\nodata	 &	RRab \\
nv23 &	-4 &	8 &	0.5511 &	2451379.821 &	0.32 &\nodata  &16.964 &	16.960	  &\nodata  &\nodata 	 &\nodata	 &	RRab \\
nv24 &	-5 &	-11 &	0.6468 &	2451379.707 &	0.35 &\nodata  &17.154 &	17.150	  &\nodata  &\nodata 	 &\nodata	 &	RRab \\
nv25 &	-1 &	-13 &	0.5967 &	2451379.769 &\nodata &\nodata  &\nodata	 & \nodata	  &\nodata  &\nodata	 &\nodata	 &	RRab \\
nv26 &	0 &	3 &	0.2960 &	2451379.662 &\nodata &\nodata  &\nodata	 &\nodata	 &\nodata  &\nodata	 &\nodata	 &	RRc \\
nv27 &	42 &	-71 &	0.6630 &	2451379.700 &	0.34 &	0.25 &	18.351 &	18.345 &	17.812 &	17.809 &	0.539 &	RRab \\
nv28 &	15 &	-30 &	0.2511 &	2451379.876 &	0.18 &	0.13 &	17.991 &	17.989 &	17.667 &	17.666 &	0.324 &	RRc \\
nv29 &	-11 &	6 &	0.3415 &	2451379.587 &\nodata &\nodata &	\nodata	 &\nodata	 &\nodata &\nodata	 &\nodata	 &	RRc? \\
nv30 &	-12 &	-6 &	0.5642 &	2451379.390 &\nodata &\nodata &	\nodata	 &\nodata	 &\nodata &\nodata	&\nodata	 &	? \\
\enddata
\label{Table1}
\end{deluxetable}

\begin{deluxetable}{lcccccc} 
\tablenum{2}
\tablewidth{0pc}
\footnotesize
\tablecaption{Fourier Coefficients: RRc Stars}
\tablehead{
  \colhead{ID} & 
  \colhead{$A_{21}$} & 
  \colhead{$A_{31}$} & 
  \colhead{$A_{41}$} & 
  \colhead{$\phi_{21}$} & 
  \colhead{$\phi_{31}$} & 
  \colhead{$\phi_{41}$} 
}
\startdata
V4   & 0.153  & 0.090  & 0.067  & 4.691  & $3.066\pm 0.085$      & 1.705  \\
V9   & 0.106  & 0.057  & 0.047  & 5.203  & $4.169\pm 0.117$      & 2.740  \\
V11  & 0.149  & 0.013  & 0.018  & 4.611  & $5.965\pm 1.218$      & 2.656  \\
V12  & 0.207: & 0.071: & 0.030: & 4.689: & $2.238\!:\pm 0.249$   & 0.962: \\
NV4  & 0.150  & 0.055  & 0.028  & 4.689  & $3.134\pm 0.287$      & 1.867  \\
NV8  & 0.044  & 0.072  & 0.038  & 5.306  & $4.456\pm 0.152$      & 3.428  \\
NV11 & 0.140  & 0.082  & 0.073  & 5.042  & $3.359\pm 0.143$      & 2.062  \\
NV12 & 0.075  & 0.051  & 0.030  & 5.384  & $4.554\pm 0.192$      & 3.348  \\
NV14 & 0.173  & 0.070  & 0.049  & 4.949  & $2.705\pm 0.158$      & 1.575  \\
NV19 & 0.058: & 0.098: & 0.055: & 4.733: & $4.511\!:\pm 0.167$   & 3.179: \\
NV26 & \nodata & \nodata & \nodata & \nodata & \nodata & \nodata  \\
NV28 & 0.012:: & 0.058:: & 0.023:: & 2.165:: & $0.604::\pm 1.037$     & 5.298:: \\
\enddata
\label{FourC}
\end{deluxetable}

\begin{deluxetable}{lllllllc} 
\tablenum{3}
\tablewidth{0pc}
\footnotesize
\tablecaption{Fourier Coefficients: RRab Stars}
\tablehead{
  \colhead{ID} & 
  \colhead{$A_{21}$} & 
  \colhead{$A_{31}$} & 
  \colhead{$A_{41}$} & 
  \colhead{$\phi_{21}$} & 
  \colhead{$\phi_{31}$} & 
  \colhead{$\phi_{41}$} & 
  \colhead{$D_m$} 
}
\startdata
V1   & 0.486   & 0.290   & 0.163   & 2.586   & 5.494   & 2.139   & 161.80   \\
V3   & 0.457   & 0.191   & 0.047   & 2.975   & 0.038   & 1.796   & 117.91   \\
V8   & 0.376   & 0.157   & 0.042   & 2.648   & 5.749   & 3.127   & 118.16   \\
V13  & 1.089:: & 1.201:: & 0.846:: & 2.706:: & 4.405:: & 5.832:: & 160.72:: \\
V14  & 0.500   & 0.356   & 0.230   & 2.346   & 5.185   & 1.962   &   4.97   \\
NV1  & 0.564   & 0.395   & 0.226   & 2.724   & 5.905   & 2.695   &  10.49   \\
NV2  & 0.542   & 0.347   & 0.239   & 2.461   & 5.317   & 1.850   &   2.28   \\
NV3  & 0.502   & 0.365   & 0.236   & 2.347   & 5.099   & 1.526   &   2.52   \\
NV5  & 0.511   & 0.284   & 0.176   & 2.433   & 5.122   & 1.708   &   2.73   \\
NV6  & 0.460   & 0.281   & 0.109   & 2.659   & 5.590   & 2.486   &   6.30   \\
NV7  & 0.405   & 0.201   & 0.121   & 2.405   & 4.957   & 0.909   &  44.38   \\
NV9  & 0.527   & 0.373   & 0.232   & 2.402   & 5.146   & 1.629   &   2.71   \\
NV10 & 0.921:: & 0.774   & 0.659:: & 0.911:: & 2.374:: & 3.610:: & 100.57:: \\
NV13 & 0.416   & 0.199   & 0.056   & 2.646   & 5.741   & 2.822   & 111.79   \\
NV15 & 0.220:  & 0.273:  & 0.130:  & 1.772:  & 4.171:  & 0.583:  &  48.56:  \\
NV16 & \nodata & \nodata & \nodata & \nodata & \nodata & \nodata & \nodata  \\
NV17 & 0.490   & 0.323   & 0.183   & 2.288   & 4.864   & 1.217   &   6.34   \\
NV18 & \nodata & \nodata & \nodata & \nodata & \nodata & \nodata & \nodata  \\
NV20 & \nodata & \nodata & \nodata & \nodata & \nodata & \nodata & \nodata  \\
NV21 & 0.525   & 0.326   & 0.231   & 2.202   & 4.637   & 0.899   &  40.03   \\
NV22 & \nodata & \nodata & \nodata & \nodata & \nodata & \nodata & \nodata  \\
NV23 & 0.491:  & 0.277:  & 0.113:  & 2.763:  & 5.568:  & 1.727:  &  30.52:  \\
NV24 & 0.489:  & 0.192:  & 0.117:  & 2.690:  & 5.795:  & 2.484:  & 118.74:  \\
NV25 & 0.624   & 0.335   & 0.252   & 2.611   & 5.204   & 1.847   &   8.91   \\
NV27 & 0.262   & 0.070   & 0.030   & 3.020   & 1.009   & 4.852   & 140.80   \\
\enddata
\label{FourAB}
\end{deluxetable}


\begin{deluxetable}{lcccc} 
\tablenum{4}
\tablewidth{0pc}
\footnotesize
\tablecaption{Fourier-Based Physical Parameters: RRc Stars}
\tablehead{
  \colhead{ID} & 
  \colhead{$M/M_{\odot}$} & 
  \colhead{$\log (L/L_{\odot})$} & 
  \colhead{$T_{\rm eff}$~(K)} & 
  \colhead{$y$} 
}
\startdata
V4   & 0.588 & 1.665 & 7430 & 0.286 \\ 
V9   & 0.494 & 1.693 & 7311 & 0.284 \\
NV4  & 0.547 & 1.614 & 7551 & 0.304 \\
NV8  & 0.478 & 1.712 & 7261 & 0.280 \\
NV11 & 0.554 & 1.661 & 7430 & 0.289 \\
NV12 & 0.451 & 1.676 & 7345 & 0.292 \\
NV14 & 0.622 & 1.655 & 7464 & 0.287 \\
Mean & $0.533\pm 0.023$ & $1.668\pm 0.012$ & $7399\pm 37$ & $0.289\pm 0.003$ \\
\enddata
\label{PhysC}
\end{deluxetable}

\noindent\begin{deluxetable}{lcccccccc} 
\tablenum{5}
\tablewidth{50pc}
\footnotesize
\tablecaption{Fourier-Based Physical Parameters: RRab Stars}
\tablehead{
  \colhead{ID} & 
  \colhead{[Fe/H]} & 
  \colhead{$\langle M_V \rangle$} & 
  \colhead{$\langle V-K \rangle$} & 
  \colhead{$\log T_{\rm e}^{\langle V-K \rangle}$} &  
  \colhead{$\langle \bv \rangle$} & 
  \colhead{$\log T_{\rm e}^{\langle\bv\rangle}$} & 
  \colhead{$\langle V-I \rangle$} & 
  \colhead{$\log T_{\rm e}^{\langle V-I\rangle}$} 
}
\startdata
V14  & $-1.17$ & 0.811 & 1.137 & 3.806 & 0.353 & 3.808 & 0.512 & 3.806  \\
NV2  & $-0.95$ & 0.799 & 1.068 & 3.813 & 0.335 & 3.816 & 0.489 & 3.811  \\
NV3  & $-0.95$ & 0.817 & 1.011 & 3.820 & 0.314 & 3.824 & 0.461 & 3.818  \\
NV5  & $-0.92$ & 0.832 & 1.025 & 3.818 & 0.310 & 3.825 & 0.457 & 3.819  \\
NV9  & $-1.04$ & 0.768 & 1.034 & 3.817 & 0.313 & 3.823 & 0.461 & 3.818  \\
Mean & $-1.01\pm 0.05$ & $0.805\pm 0.011$ & $1.055\pm 0.023$ & $3.815\pm 0.002$ & $0.325\pm 0.019$ & $3.819\pm 0.003$ & $0.476\pm 0.024$ & $3.814\pm 0.006$ \\
\enddata
\label{PhysAB}
\end{deluxetable}

\clearpage

\begin{deluxetable}{lccccc} 
\tablenum{6}
\tablewidth{0pc}
\footnotesize
\tablecaption{Comparison between Mean Fourier-Based Physical Parameters for M75 
and Other Globular Clusters: RRc Stars}
\tablehead{
  \colhead{Cluster} & 
  \colhead{Oo type} & 
  \colhead{${\rm [Fe/H]}$} & 
  \colhead{$\langle M/M_{\odot}\rangle$} & 
  \colhead{$\langle \log\,(L/L_{\odot}) \rangle$} & 
  \colhead{$\langle T_{\rm e}\rangle$~(K)} 
}
\startdata
NGC 6441        &  ?   & $-0.53$ & 0.47 & 1.65 & 7408 \\
NGC 6388        &  ?   & $-0.60$ & 0.48 & 1.62 & 7495 \\
NGC 6362        & OoI  & $-0.95$ & 0.53 & 1.67 & 7429 \\
NGC 6171 (M107) & OoI  & $-1.04$ & 0.54 & 1.65 & 7448 \\
NGC 5904 (M5)   & OoI  & $-1.29$ & 0.54 & 1.69 & 7353 \\
NGC 6864 (M75)  &  ?   & $-1.32$ & 0.53 & 1.67 & 7399 \\
NGC 6229        & OoI  & $-1.43$ & 0.56 & 1.69 & 7315 \\ 
NGC 6934        & OoI  & $-1.54$ & 0.63 & 1.72 & 7290 \\ 
NGC 5272 (M3)   & OoI  & $-1.57$ & 0.59 & 1.71 & 7315 \\
NGC 6809 (M55)  & OoII & $-1.81$ & 0.53 & 1.75 & 7193 \\ 
NGC 2298        & OoII & $-1.85$ & 0.59 & 1.75 & 7200 \\
NGC 4590 (M68)  & OoII & $-2.06$ & 0.71 & 1.79 & 7145 \\
NGC 7078 (M15)  & OoII & $-2.25$ & 0.73 & 1.80 & 7136 \\
\enddata
\label{GCsC}
\end{deluxetable}

\begin{deluxetable}{lcccc} 
\tablenum{7}
\tablewidth{0pc}
\footnotesize
\tablecaption{Comparison between Mean Fourier-Based Physical Parameters for M75 
and Other Globular Clusters: RRab Stars}
\tablehead{
  \colhead{Cluster} & 
  \colhead{Oo type} & 
  \colhead{${\rm [Fe/H]}$} & 
  \colhead{$\langle M_V \rangle$} & 
  \colhead{$\langle T_{\rm e}\rangle^{\langle V-K \rangle}$~(K)} 
}
\startdata
NGC 6441        &  ?   & $-0.53$ & 0.68 & 6607 \\
NGC 6388        &  ?   & $-0.60$ & 0.66 & 6607 \\
NGC 6362        & OoI  & $-0.95$ & 0.86 & 6555 \\
NGC 6171 (M107) & OoI  & $-1.04$ & 0.85 & 6619 \\
NGC 1851        & OoI  & $-1.22$ & 0.80 & 6494 \\ 
NGC 5904 (M5)   & OoI  & $-1.29$ & 0.81 & 6465 \\
NGC 6864 (M75)  &  ?   & $-1.32$ & 0.81 & 6529 \\
NGC 6229        & OoI  & $-1.43$ & 0.82 & 6478 \\ 
NGC 6934        & OoI  & $-1.54$ & 0.81 & 6455 \\ 
NGC 5272 (M3)   & OoI  & $-1.57$ & 0.78 & 6438 \\
NGC 6809 (M55)  & OoII & $-1.81$ & 0.68 & 6325 \\ 
\enddata
\label{GCsAB}
\end{deluxetable}

\end{document}